\documentclass[epj]{svjour}
\pdfoutput=1 

\usepackage{amsmath}
\usepackage{amssymb} 
\usepackage{graphicx}

\usepackage{xcolor}
\usepackage{hyperref}
\hypersetup{
    colorlinks=true,       
    linkcolor=blue,          
    citecolor=blue,        
    filecolor=blue,      
    urlcolor=blue           
}

\newcommand{\GR}{G_{\mathrm{R}}}
\newcommand{\Grr}{G_{\mathrm{rr}}}
\newcommand{\Gpr}{G_{\mathrm{pr}}}
\newcommand{\Gqr}{G_{\mathrm{qr}}}
\newcommand{\Gqrd}{G_{\mathrm{qrd}}}
\newcommand{\Gqrnd}{G_{\mathrm{qrnd}}}
\newcommand{\WR}{W_{\mathrm{R}}}
\newcommand{\Wrr}{W_{\mathrm{rr}}}
\newcommand{\Wpr}{W_{\mathrm{pr}}}
\newcommand{\Wqr}{W_{\mathrm{qr}}}
\newcommand{\Wqrnd}{W_{\mathrm{qrnd}}}

\begin{document}

\title{Influence of petroleum and gas trade on EU economies
from the reduced Google matrix analysis of UN COMTRADE data} 

\author{C\'elestin Coquid\'e$^1$, Leonardo Ermann$^2$, Jos\'e Lages$^1$ 
\and D.L.Shepelyansky$^{3}$}

\institute{
Institut UTINAM, OSU THETA, 
Universit\'e de Bourgogne Franche-Comt\'e, CNRS, Besan\c con, France
\and
Departamento de F\'isica Te\'orica, GIyA, CNEA, Av. Libertador 8250, 
(C1429BNP) Buenos Aires, Argentina.
\and
Laboratoire de Physique Th\'eorique, IRSAMC, 
Universit\'e de Toulouse, CNRS, UPS, 31062 Toulouse, France
}

\titlerunning{Influence of petroleum and gas trade on EU economies}
\authorrunning{C.~Coquid\'e {\it et al.}}

\abstract{
Using the United Nations COMTRADE database \cite{comtrade} 
we apply the reduced Google matrix (REGOMAX) algorithm  
to analyze  the multiproduct world trade 
in years 2004-2016.
Our approach allows to determine the trade balance sensitivity of a group of countries to a specific product price increase from a specific exporting country 
taking into account all direct and indirect trade pathways 
via all world countries exchanging 61 UN COMTRADE identified trade products.
On the basis of this approach 
we present the influence of trade in petroleum and gas products
from Russia, USA, Saudi Arabia and Norway
determining the sensitivity of each EU country.
We show that the REGOMAX approach provides
a new and more detailed analysis
of trade influence propagation
comparing to the usual approach based on export and
import flows.
}



\date{Dated:  February 5, 2019}

\maketitle

\section{Introduction}

The statistical data of UN COMTRADE \cite{comtrade}
and the World Trade Organization (WTO) Statistical Review 2018 \cite{wto2018}
demonstrate the vital importance of 
the international trade between world countries 
for their development and pro\-gress.
Also the whole world economy deeply depends on the world trade \cite{krugman2011}. 
At present the UN COMTRADE data\-base contains data for $N_c=294$ UN countries
with up to $N_p \approx 10^4$ trade products.
Thus the whole matrix of trade monetary flows
reaches a large size $N=N_p N_c \sim 10^6$.
In fact for each year the commercial exchange between countries
represents the directed network with
transactions of various commodities (products)
expressed in their US dollar (USD) values of the given year.

It is clear that the recent research developments
in the field of complex networks (see e.g. \cite{dorogovtsev})
should find useful applications for analysis of this
multiproduct World Trade Network (WTN).
In \cite{wtn1,wtn2} it was proposed to use the 
methods of the Google matrix $G$, PageRank and CheiRank algorithms
for analysis of the WTN. The PageRank algorithm
had been invented by Brin and Page \cite{brin}
for the ranking of nodes of the World Wide Web (WWW)
being at the foundation grounds of the Google search engine
\cite{brin,meyer}. The applications of these methods
to a variety of real directed networks are described 
in \cite{rmp2015}. In contrast to the usual economy approach based on
bilateral import and export flows,
the Google matrix analysis 
treats all world countries on equal grounds
(since all columns with outgoing country flows of $G$
are normalized to unity so that rich and poor 
countries have equal consideration)
and also the PageRank and CheiRank algorithms take into account 
the whole chain of transactions incorporating 
the importance of specific 
network nodes. This is drastically different 
from the simple bilateral transactions of import and export. 

Usually in directed networks, like WWW or Wikipedia,
the PageRank vector of the Google matrix
plays the dominant role since its components are on average proportional
to the number of ingoing links.
For the  WTN the ingoing  flows are related to import.
However, the outgoing flows, related 
to export, are also important for trade.
Thus we also use the Google matrix $G^*$, constructed from the inverted 
transaction flows, with its PageRank eigenvector,
called CheiRank vector \cite{linux,wikizzs}.
The components of this vector are on average proportional 
to the number of outgoing links in the original WTN. 
The construction rules 
of $G$ and $G^*$ for the case of multiproduct WTN
are described in detail in \cite{wtn2}.

In many cases it is important to know the effective interactions
of trade transactions for a specific region (i.e., for selected nodes of the global network)
on which one wants to focus the analysis.
This requires to know not only direct links between nodes
but also the indirect (or hidden) links
which connect the selected nodes via the remaining part of the global network.
Recently the reduced Google matrix (REGOMAX) algorithm 
has been invented in \cite{greduced} and tested with 
various directed networks of Wikipedia \cite{politwiki,wrwu2017}
and protein-protein interactions \cite{proteinplos}
showing its efficiency.
This algorithm originates from the scattering theory of nuclear and 
mesoscopic physics and the field of quantum chaos.
In this work, using the COMTRADE data, 
we apply the REGOMAX algorithm to analyze the influence
on European Union (EU) countries 
of petroleum and gas trade from Russia (RU), USA (US), Saudi Arabia (SA) 
and Norway (NO). With this approach we are able to measure 
the sensitivity of EU countries 
to the supply of petroleum and gas from one of these four countries 
taking into account the global WTN, i.e., taking into account all direct 
and indirect transactions of 61 major products with the rest of the world.

We note that there is a variety of papers with network methods
applied to financial and trade networks 
(see e.g. \cite{vespignani,fagiolo1,hedeem,fagiolo2,garlaschelli2010,japan1}).
However, the applications of the PageRank algorithm
to the WTN is rarely used (see e.g. one of the first cases in 
\cite{benedictis}) but the outgoing flows with the CheiRank
analysis were not considered apart from \cite{wtn1,wtn2}.
The analysis of hubs and authorities is performed in \cite{plosjapan}
but in our opinion this approach has lower performance comparing to the
Google matrix methods. Thus for the bitcoin transaction network the 
Google matrix methods demonstrate the existence of oligarchy type structure
\cite{bitcoin}.
Till present the matrix methods
are rather rarely used in the field of transactions
even if it was shown that the Random Matrix Theory
finds useful applications for financial and credit risk analysis
\cite{bouchaud,guhr}. 
The methods of statistical mechanics also demonstrated their efficiency
for analysis of market economies \cite{marsili}.
However, the flows considered in \cite{bouchaud,guhr}
are non-directional while the WTN typically describes directed flows.
Due to these reasons we hope that the REGOMAX algorithm will 
find further useful application for the treatment of
trade and financial transactions. 

The paper is constructed as follows: in Section 2, we construct the Google matrix for the World Trade Network and introduce the REGOMAX method. In Section 3, we present the network structure of petroleum and gas trade in EU exhibiting direct and indirect effects of petroleum and gas trade between EU economies and non EU major actors as Russia, Saudi Arabia and USA. We also investigate the EU countries trade balance sensitivity to Russian, Saudi Arabian, and US petroleum and to Russian and Norwegian gas over the time period 2004-2016.

\section{Methods}
\label{sec:2}

We collected the multiproduct (multicommodities)
trade data from UN COMTRADE database \cite{comtrade}
for $N_c=227$ countries, $N_p=61$ products given by
2 digits from the
Standard International Trade Classification (SITC) Rev. 1, and
for years $2004$, $2008$, $2012$, $2016$.
Following the approach developed in \cite{wtn2},
for a given year, we build $N_p$ money matrices $M^p_{c,c^\prime}$ of 
the WTN defined as
\begin{equation}
M^p_{c,c^\prime}=
\begin{array}{c}
\text{product $p$ transfer (in USD)}\\
\text{from country $c^\prime$ to $c$}
\end{array}
\label{eq1}
\end{equation}
with country indexes $c,c^\prime=1,\ldots,N_c$ and 
product index $p=1,\ldots,N_p$. 
For future notation we also define
\begin{equation}
V^p_c=\sum_{c^\prime} M^p_{c,c^\prime} \, , \,\;
V^{*p}_c=\sum_{c^\prime} M^p_{c^\prime,c} .
\label{eq2}
\end{equation}
which are the volume of imports ($V_c^p$) and exports ($V_c^{*p}$)
for a given country $c$ and a given product $p$.
The global import and export volumes are given by $V_c=\sum_p V^p_c$ and $V^*_c = \sum_p V^{*p}_c$.
Thus the
ImportRank ($\hat{P}$) and ExportRank ($\hat{P}^*$) vector probabilities
are given by the normalized import and export volumes
\begin{equation}
\hat{P}_{i} = {V^p_c}/{V} \, , \,\;
\hat{P}^*_{i} = {V^{*p}_c}/{V} \, ,
\label{eq3}
\end{equation}
where $i=p+(c-1)N_p\in\{1,\dots,N=N_cN_p\}$ is the index associated to the country 
$c$ -- product $p$ couple, and the total trade volume is
$V=\sum_{p,c,c^\prime} M^p_{c,c^\prime}=\sum_{p,c}V^p_c=\sum_{p,c}V^{*p}_c$.

The list of 61 products and 227 countries are given in \cite{wtn2}.

\subsection{Google matrix construction for the WTN}

The Google matrices $G$ for the direct trade flow
and $G^*$ for the inverted trade flow have the size $N=N_c N_p = 227 \times 61 = 13847$ 
and are constructed as it is described in \cite{wtn2}.
By the definition the sum of elements in each column
is equal to unity. The Google matrices have the form
\begin{equation}
\begin{array}{ccc}
G_{ij}&=& \alpha S_{ij}+(1-\alpha) v_i,\\
G^*_{ij}&=&\alpha S^*_{ij}+(1-\alpha) v^*_i,
\end{array}
\label{eq4}
\end{equation}
where $\alpha \in ]0,1]$ is the damping factor, 
and $v_i$ and $v_i^*$ are components of positive column vectors called \emph{personalization vectors} 
with $\sum_i v_i=\sum_i v_i^*=1$ \cite{meyer}. In this work we fix $\alpha=0.5$,
its variation in the range $[0.5,0.9]$ does not significantly affect
the results. The PageRank $P$ and CheiRank $P^*$ vectors 
have each an eigenvalue $\lambda=1$ since $GP =P$ and $G^*P^*=P^*$.
According to the Perron-Frobenius theorem
the components $\{P_i\}_{i=1,\dots,N}$ and $\{P^*_i\}_{i=1,\dots,N}$ are positive
and give probabilities to find
a random surfer (seller) traveling on the network of $N$ nodes.
The PageRank $K$ and CheiRank $K^*$ indexes
are defined from the decreasing ordering of probabilities of PageRank vector $P$ and of CheiRank vector $P^*$ as
$P(K)\ge P(K+1)$ and $P^*(K^*)\ge P^*(K^*+1)$ with $K,K^*=1,\ldots,N$.
A similar definition of ranks 
from import and export trade volume can be also done 
via probabilities
$\hat{P}_p$, $\hat{P}^*_p$, $\hat{P}_c$, $\hat{P}^*_c$,
$\hat{P}_{pc}$, $\hat{P}^*_{pc}$ and 
corresponding indexes
$\hat{K}_p$, $\hat{K}^*_p$, $\hat{K}_c$, $\hat{K}^*_c$, $\hat{K}$, $\hat{K}^*$.

The matrices $S$ and $S^*$  are built from money matrices $M^p_{c,c'}$ as
\begin{equation}
\begin{array}{l}
S_{ii^\prime}=\left\{\begin{array}{cl}   
M^p_{c,c^\prime} \delta_{pp^\prime}/V^{*p}_{c^\prime}& 
\text{    if } V^{*p^\prime}_{c^\prime}\ne0\\ 
1/N & \text{    if } V^{*p\prime}_{c^\prime}=0\\ 
\end{array}\right.\\
\\
S^*_{ii^\prime}=\left\{\begin{array}{cl}   
M^p_{c^\prime,c} \delta_{pp^\prime}/V^{p}_{c^\prime}& 
\text{    if } V^{p^\prime}_{c^\prime}\ne0\\ 
1/N & \text{    if } V^{p^\prime}_{c^\prime}=0\\ 
\end{array}\right.
\label{eq5}
\end{array}
\end{equation}
where $c,c^\prime=1,\ldots,N_c$; $p,p^\prime=1,\ldots,N_p$; 
$i=p+(c-1)N_p$; $i^\prime=p^\prime+(c^\prime-1)N_p$; 
and therefore $i,i^\prime=1,\ldots,N$.

Following \cite{wtn2} we defined the personalized vectors in (\ref{eq4})
via the relative import and export product volume per country
\begin{equation}
v_i = \frac{V^p_c}{N_c \sum_{p^\prime} V^{p^\prime}_c} \, , \;
v^*_i = \frac{V^{*p}_c}{N_c \sum_{p^\prime} V^{*p^\prime}_c} \, ,
\label{eq6}
\end{equation}
using the definitions (\ref{eq2}) and the relation
$i=p+(c-1)N_p$. In this way we obtain the first iteration 
for PageRank $P$ and CheiRank $P^*$ vectors
keeping the democracy in countries and 
proportionality of products to their trade volume.
Then in the second iteration  we use  the personalized
vectors from the results of the first iteration
\begin{equation}
v^\prime_i = \frac{P_p^{\phantom{*}}}{N_c} \, , \;
v^{\prime *}_i = \frac{P^*_p}{N_c} \, .
\label{eq7} 
\end{equation}
Here we use the tracing over product or countries getting respectively
$P_c=\sum_{p} P_{pc}=\sum_{p}P\left(p+(c-1)N_p\right)$ 
and $P^*_c= \sum_{p} P^*_{pc}=\sum_{p}P^*\left(p+(c-1)N_p\right)$ 
with their corresponding $K_c$ and $K^*_c$ indexes.
Also after tracing over countries we obtain
$P_p=\sum_{c} P_{pc} =\sum_{c}P\left(p+(c-1)N_p\right)$  and 
$P^*_p=\sum_{c} P^*_{pc}=\sum_{c}P^*\left(p+(c-1)N_p\right)$ 
with their corresponding product indexes $K_p$ and $K^*_p$
($P_p$, ${P^*}_p$ are used in (\ref{eq7})).
This second iteration is used for further construction
of $G$ and $G^*$ matrices with which we work
in the following.

\subsection{Reduced Google matrix for the WTN}

The REGOMAX algorithm, invented in \cite{greduced}, 
is described in detail in \cite{politwiki}.
Here we give the main elements of this method
keeping the notations of \cite{politwiki}.

The reduced Google matrix $\GR$ is constructed for a selected subset of
$N_r$ nodes. 
The construction is based on concepts of scattering theory 
used in different fields in\-clu\-ding mesoscopic and nuclear physics, and  
quantum chaos. It captures, in a $N_r$-by-$N_r$ Perron-Frobenius matrix,
the full contribution of direct and indirect interactions happening 
in the full $G$ matrix  between  $N_r$ selected nodes of interest. 
In addition the PageRank probabilities of the $N_r$ nodes are the same 
as for the global network with $N$ nodes,
up to a constant factor taking into account that 
the sum of PageRank probabilities over $N_r$
nodes is unity. The $(i,j)$-element of $\GR$ 
can be interpreted as the probability for a random surfer starting at 
node $j$ to arrive in node $i$ using direct and indirect interactions. 
Indirect interactions refer to pathways composed in part of nodes different 
from the $N_r$ ones of interest.    
The intermediate computation steps of $\GR$ offer 
a decomposition of $\GR$ into matrices that clearly distinguish 
direct from indirect interactions: 
$\GR = \Grr + \Gpr + \Gqr$ \cite{politwiki}.
Here $\Grr$ is given by the direct links between selected 
$N_r$ nodes in the global $G$ matrix with $N$ nodes, 
 $\Gpr$ is usually rather close to 
the matrix in which each column is given by 
the PageRank vector $P_r$. 
Due to that $\Gpr$ does not provide much information about direct 
and indirect links between selected nodes.
The interesting role is played by $\Gqr$, which takes 
into account all indirect links between
selected nodes appearing due to multiple pathways via 
the $N$ global network nodes (see~\cite{greduced,politwiki}).
The matrix  $\Gqr = \Gqrd + \Gqrnd$ has diagonal ($\Gqrd$)
and non-diagonal ($\Gqrnd$) parts where $\Gqrnd$
 describes indirect interactions between nodes.
The explicit formulas with the mathematical and numerical computation 
methods of all three matrix components of $\GR$ are given 
in \cite{greduced,politwiki}. 
We discuss the properties of these matrix components below
for the multiproduct WTN. 

\subsection{WTN datasets}

With the REGOMAX approach we consider
$27$ EU countries dated by 2008 and presented in Table~\ref{tab:petroleum16}
and Table~\ref{tab:gas16}; countries are
marked by their country code ISO 3166-1 alpha-2 \cite{wikicc}.
The Table of $61$ products is given in \cite{wtn2}.

In Table~\ref{tab:petroleum16} in addition to 27 EU countries (marked by blue)
we also take $10$ best non-EU petroleum (SITC Rev.1 code $p=33$ 
for petroleum and petroleum products) exporters in 2016
(marked by red) showing their
 PageRank, CheiRank, ImportRank and ExportRank in 2016.
Here the PageRank and CheiRank are given by the local order
of $P_{pc}$ and $P^*_{pc}$ with fixed $p=33$ with highest probabilities
at index being $1,2,...$ (probability in decreasing order).
In the same way  ImportRank and ExportRank are
obtained from $\hat{P}_{pc}$ and ${\hat{P}^*}_{pc}$
at fixed $p=33$.

For petroleum we see in Table~\ref{tab:petroleum16} 
that in 2016 the top position is 
taken by Russia in  CheiRank and ExportRank
while USA is the first in PageRank and ImportRank. 
We also see that for CheiRank not only the trade volume
counts but also the broad trade network of a given country.
Thus Saudi Arabia (SA) is 2nd in ExportRank but it has only 6th position 
in CheiRank since its trade is mainly oriented towards US.
Another example is Singapore (SG) which goes from 4th position in ImportRank
to 2nd position in PageRank demonstrating the importance of broad trade 
relations of SG. Among EU countries the first place in all 4 ranks is
taken by Netherlands (NL) due to its broad commercial maritime connections. 

For gas in Table~\ref{tab:gas16} we have similar observations. Although France (FR), Italy (IT) and UK (GB) occupy the first ImportRank places for EU countries, i.e., they are the top EU importer by volume trade of gas, NL and Belgium (BE) supersede them in PageRank top positions, indicating that NL and BE import gas from more diverse and important sources than FR, IT and GB.
Also Qatar (QA) is first in ExportRank but is only at the 4th position 
in CheiRank due to its rather specific trade orientation.

\begin{table}[!th]
\begin{center}
\caption{List of 27 EU countries (in blue)  and 10 best non-EU exporters 
regarding to ExportRank for SITC Rev.1 code $p=33$ 
(petroleum and petroleum products, in red), 
sorted by
PageRank, 
CheiRank,
ImportRank
and
ExportRank
order from UN COMTRADE 2016.}
\setlength{\tabcolsep}{3pt}
\begin{tabular}{rcccc}
&\small PageRank&\small CheiRank&\small ImportRank&\small ExportRank\\
\hline
1&{\color{red}US}&{\color{red}RU}&{\color{red}US}&{\color{red}RU}\\
2&{\color{red}SG}&{\color{red}US}&{\color{blue}NL}&{\color{red}SA}\\
3&{\color{blue}NL}&{\color{red}AE}&{\color{red}IN}&{\color{red}US}\\
4&{\color{red}IN}&{\color{red}IN}&{\color{red}SG}&{\color{red}AE}\\
5&{\color{blue}FR}&{\color{red}SG}&{\color{blue}DE}&{\color{blue}NL}\\
6&{\color{blue}DE}&{\color{red}SA}&{\color{blue}IT}&{\color{red}CA}\\
7&{\color{blue}ES}&{\color{blue}NL}&{\color{blue}FR}&{\color{red}IQ}\\
8&{\color{blue}GB}&{\color{blue}BE}&{\color{blue}GB}&{\color{red}SG}\\
9&{\color{blue}IT}&{\color{blue}GR}&{\color{blue}BE}&{\color{red}KW}\\
10&{\color{blue}BE}&{\color{red}NG}&{\color{blue}ES}&{\color{red}NG}\\
11&{\color{red}CA}&{\color{blue}IT}&{\color{red}CA}&{\color{red}IN}\\
12&{\color{red}AE}&{\color{blue}DE}&{\color{blue}SE}&{\color{blue}GB}\\
13&{\color{red}NG}&{\color{red}CA}&{\color{blue}PL}&{\color{blue}BE}\\
14&{\color{blue}PL}&{\color{red}IQ}&{\color{red}NG}&{\color{blue}DE}\\
15&{\color{blue}SI}&{\color{red}KW}&{\color{red}AE}&{\color{blue}IT}\\
16&{\color{blue}CZ}&{\color{blue}GB}&{\color{blue}GR}&{\color{blue}ES}\\
17&{\color{blue}AT}&{\color{blue}ES}&{\color{blue}FI}&{\color{blue}FR}\\
18&{\color{blue}SE}&{\color{blue}FR}&{\color{blue}AT}&{\color{blue}GR}\\
19&{\color{blue}HU}&{\color{blue}FI}&{\color{blue}PT}&{\color{blue}SE}\\
20&{\color{blue}PT}&{\color{blue}SE}&{\color{blue}LV}&{\color{blue}FI}\\
21&{\color{blue}RO}&{\color{blue}PT}&{\color{blue}MT}&{\color{blue}LT}\\
22&{\color{blue}BG}&{\color{blue}RO}&{\color{blue}CZ}&{\color{blue}DK}\\
23&{\color{blue}SK}&{\color{blue}DK}&{\color{blue}DK}&{\color{blue}PL}\\
24&{\color{blue}GR}&{\color{blue}BG}&{\color{blue}LT}&{\color{blue}PT}\\
25&{\color{blue}MT}&{\color{blue}LT}&{\color{blue}RO}&{\color{blue}RO}\\
26&{\color{red}SA}&{\color{blue}PL}&{\color{blue}IE}&{\color{blue}BG}\\
27&{\color{red}RU}&{\color{blue}HU}&{\color{blue}HU}&{\color{blue}SK}\\
28&{\color{blue}LT}&{\color{blue}AT}&{\color{blue}SK}&{\color{blue}AT}\\
29&{\color{blue}IE}&{\color{blue}SK}&{\color{red}SA}&{\color{blue}LV}\\
30&{\color{blue}CY}&{\color{blue}LV}&{\color{blue}BG}&{\color{blue}MT}\\
31&{\color{blue}DK}&{\color{blue}MT}&{\color{blue}SI}&{\color{blue}HU}\\
32&{\color{blue}FI}&{\color{blue}CZ}&{\color{red}RU}&{\color{blue}CZ}\\
33&{\color{blue}LV}&{\color{blue}SI}&{\color{blue}EE}&{\color{blue}EE}\\
34&{\color{blue}LU}&{\color{blue}CY}&{\color{blue}LU}&{\color{blue}SI}\\
35&{\color{red}IQ}&{\color{blue}EE}&{\color{blue}CY}&{\color{blue}IE}\\
36&{\color{blue}EE}&{\color{blue}IE}&{\color{red}IQ}&{\color{blue}CY}\\
37&{\color{red}KW}&{\color{blue}LU}&{\color{red}KW}&{\color{blue}LU}\\
 \hline
 \end{tabular}
 \label{tab:petroleum16}
 \end{center}
\end{table}

\begin{table}[!th]
\begin{center}
\caption{List of 27 EU countries (in blue)  and 10 best non-EU exporters 
regarding to ExportRank for SITC Rev.1 code $p=34$ 
(gas, natural and manufactured, in red), 
sorted by
PageRank,
CheiRank,
ImportRank
and
ExportRank
order from UN COMTRADE 2016.
}
\setlength{\tabcolsep}{3pt}
\begin{tabular}{rcccc}
&\small PageRank&\small CheiRank&\small ImportRank&\small ExportRank\\
\hline
1&{\color{blue}NL}&{\color{red}US}&{\color{blue}FR}&{\color{red}QA}\\
2&{\color{blue}BE}&{\color{red}CA}&{\color{blue}IT}&{\color{red}NO}\\
3&{\color{blue}FR}&{\color{red}RU}&{\color{blue}GB}&{\color{red}RU}\\
4&{\color{blue}IT}&{\color{red}QA}&{\color{red}US}&{\color{red}US}\\
5&{\color{blue}GB}&{\color{red}NO}&{\color{blue}DE}&{\color{red}AU}\\
6&{\color{blue}ES}&{\color{red}AU}&{\color{blue}BE}&{\color{red}DZ}\\
7&{\color{blue}HU}&{\color{blue}NL}&{\color{blue}ES}&{\color{red}MY}\\
8&{\color{red}US}&{\color{blue}GB}&{\color{blue}NL}&{\color{blue}BE}\\
9&{\color{blue}DE}&{\color{red}DZ}&{\color{red}AE}&{\color{red}CA}\\
10&{\color{blue}PT}&{\color{red}AE}&{\color{red}CA}&{\color{red}AE}\\
11&{\color{blue}BG}&{\color{blue}BE}&{\color{red}ID}&{\color{red}ID}\\
12&{\color{blue}SK}&{\color{blue}DE}&{\color{blue}CZ}&{\color{blue}NL}\\
13&{\color{blue}PL}&{\color{blue}IT}&{\color{blue}SK}&{\color{blue}GB}\\
14&{\color{blue}SI}&{\color{blue}FR}&{\color{blue}PT}&{\color{blue}DE}\\
15&{\color{red}CA}&{\color{blue}SE}&{\color{blue}HU}&{\color{blue}FR}\\
16&{\color{blue}RO}&{\color{red}ID}&{\color{blue}PL}&{\color{blue}ES}\\
17&{\color{red}ID}&{\color{blue}DK}&{\color{red}MY}&{\color{blue}AT}\\
18&{\color{blue}GR}&{\color{red}MY}&{\color{blue}IE}&{\color{blue}SK}\\
19&{\color{red}RU}&{\color{blue}GR}&{\color{blue}GR}&{\color{blue}CZ}\\
20&{\color{red}MY}&{\color{blue}PL}&{\color{blue}SE}&{\color{blue}IT}\\
21&{\color{blue}CZ}&{\color{blue}ES}&{\color{blue}BG}&{\color{blue}PL}\\
22&{\color{blue}SE}&{\color{blue}AT}&{\color{blue}LT}&{\color{blue}SE}\\
23&{\color{red}AU}&{\color{blue}PT}&{\color{blue}RO}&{\color{blue}HU}\\
24&{\color{blue}IE}&{\color{blue}HU}&{\color{blue}LV}&{\color{blue}DK}\\
25&{\color{red}AE}&{\color{blue}IE}&{\color{blue}SI}&{\color{blue}RO}\\
26&{\color{blue}LT}&{\color{blue}SK}&{\color{red}AU}&{\color{blue}SI}\\
27&{\color{red}NO}&{\color{blue}LT}&{\color{red}RU}&{\color{blue}PT}\\
28&{\color{blue}AT}&{\color{blue}RO}&{\color{blue}DK}&{\color{blue}GR}\\
29&{\color{blue}DK}&{\color{blue}CZ}&{\color{blue}EE}&{\color{blue}LT}\\
30&{\color{blue}CY}&{\color{blue}SI}&{\color{red}NO}&{\color{blue}LU}\\
31&{\color{blue}EE}&{\color{blue}LV}&{\color{blue}LU}&{\color{blue}LV}\\
32&{\color{blue}MT}&{\color{blue}BG}&{\color{blue}FI}&{\color{blue}FI}\\
33&{\color{blue}LV}&{\color{blue}FI}&{\color{blue}AT}&{\color{blue}MT}\\
34&{\color{blue}LU}&{\color{blue}LU}&{\color{blue}CY}&{\color{blue}IE}\\
35&{\color{blue}FI}&{\color{blue}MT}&{\color{blue}MT}&{\color{blue}EE}\\
36&{\color{red}QA}&{\color{blue}EE}&{\color{red}QA}&{\color{blue}BG}\\
37&{\color{red}DZ}&{\color{blue}CY}&{\color{red}DZ}&{\color{blue}CY}\\
 \hline
 \end{tabular}
\label{tab:gas16}
\end{center}
\end{table}

\subsection{Sensitivity of trade balance}
\label{subs:sensitivity}

As in \cite{wtn2},
we determine the trade balance of a given country
with PageRank and CheiRank probabilities
as $B_c = (P^*_c - P_c)/(P^*_c + P_c)$
and in a similar way via ImportRank and ExportRank probabilities
as $\hat{B}_c=  ({\hat{P}^*}_c - \hat{P}_c)/({\hat{P}^*}_c + \hat{P}_c)$.
The sensitivity of 
trade balance $B_c$ to the price of petroleum or gas can be obtained 
by the change of the corresponding money volume
flow related to code $33$ or $34$ by multiplying it by $(1+\delta)$,
computing all rank probabilities and then
the derivative $dB_c/d\delta$. 

This approach was used in \cite{wtn2}.
However, in this way there we had the effect of  global price change
of petroleum or gas for all countries.
Here, we want to determine the sensitivity of country balance
to a flow of petroleum from a specific country (e.g. RU, US, or SA).
Thus we first compute all 4 matrix components of the reduced Google matrix
$\GR$, $\Gpr$, $\Grr$, $\Gqr$ and then we recompute these matrices
with the price modification factor $(1+\delta)$ applied
only for the trade of a given EU country with 
Russia (or with US, or SA).

\begin{figure*}[!th]
\centering
\includegraphics[width=\columnwidth]{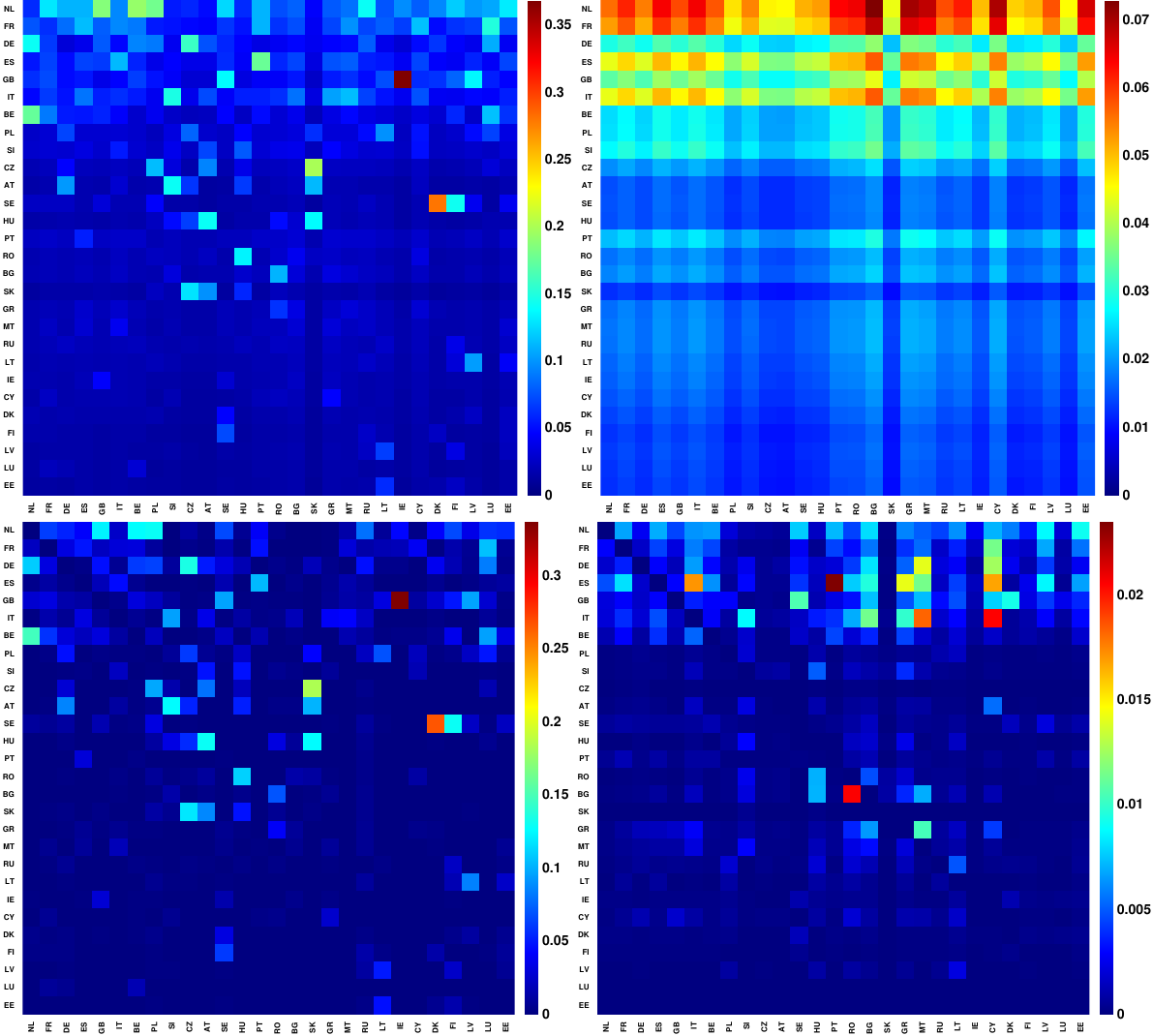}\hfill
\includegraphics[width=\columnwidth]{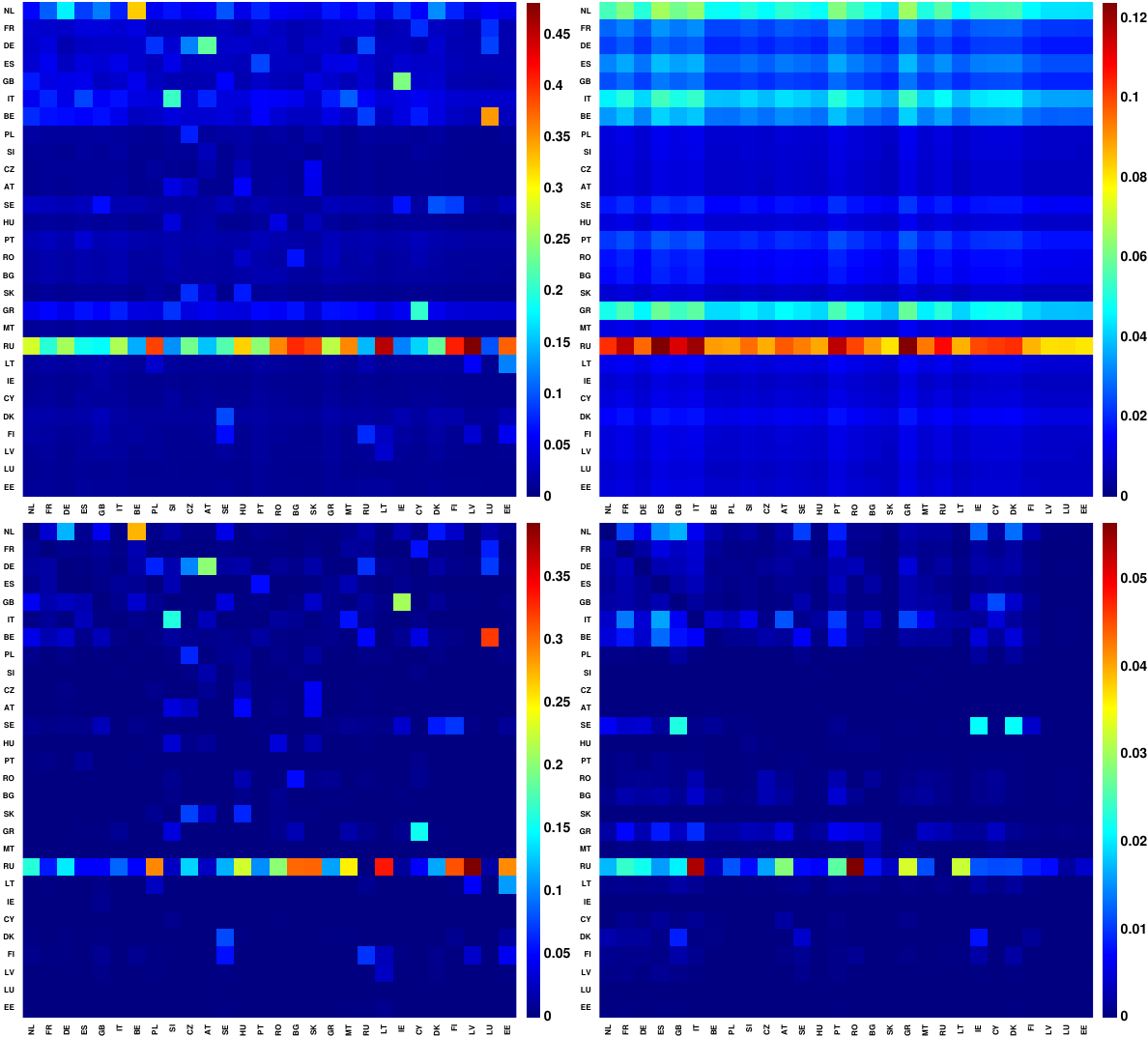}
\caption{
Left four panels: reduced Google matrix $\GR$ (top left) and its matrix components 
$\Gpr$ (top right), $\Grr$ (bottom left) and $\Gqrnd$ (bottom right) 
for the petroleum product (code $p=33$) exchanged among the 27 EU countries 
and Russia in 2016. Right four panels: the same as on the left but
for reduced Google matrix $\GR^*$ and its three matrix components
in the same order as on the left.
Here, the EU countries and RU are ordered as in
the PageRank column of Table~\ref{tab:petroleum16}.
}
\label{fig1}
\end{figure*}

\begin{figure*}[h]
\centering
\includegraphics[width=\columnwidth]{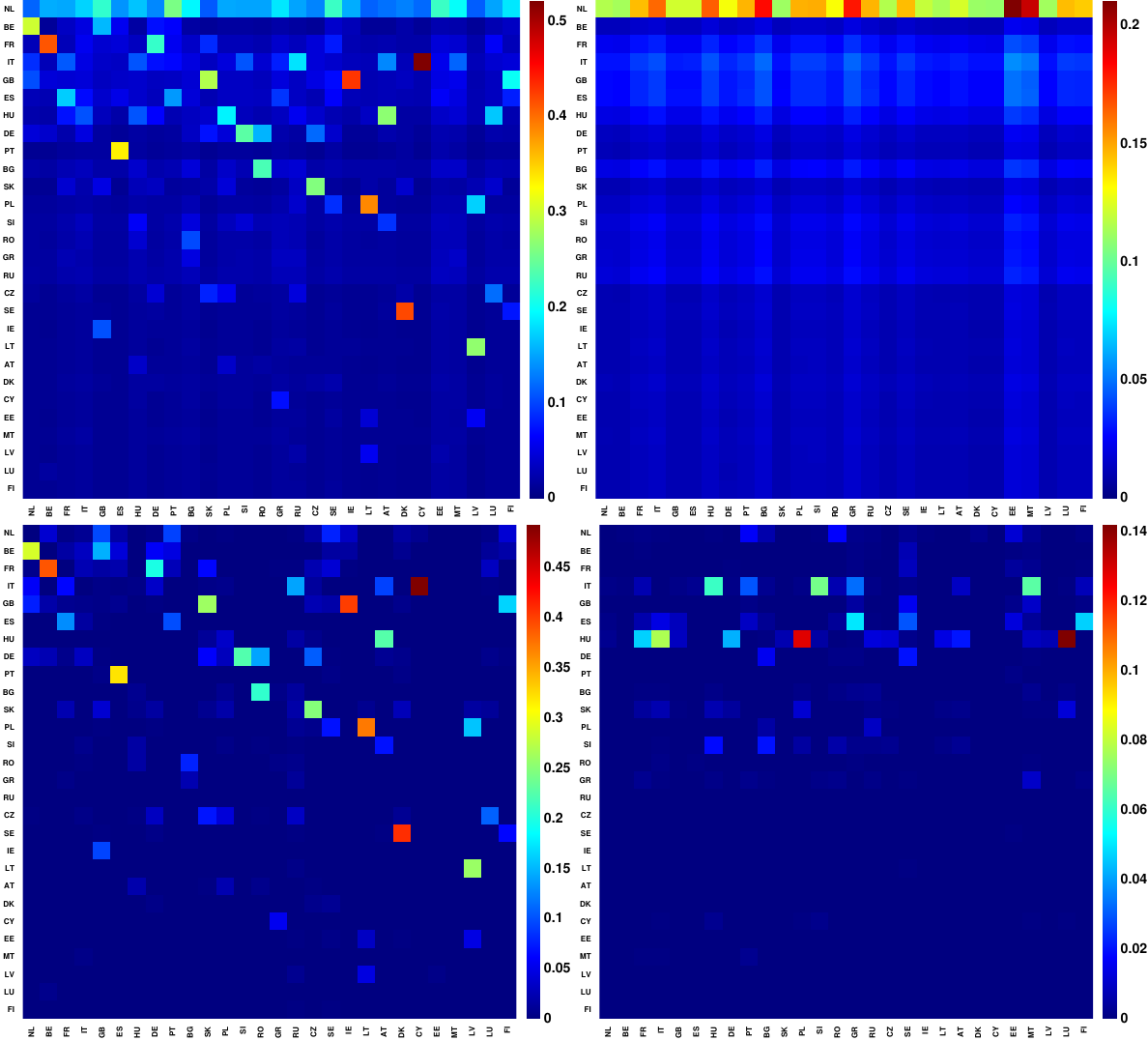}\hfill
\includegraphics[width=\columnwidth]{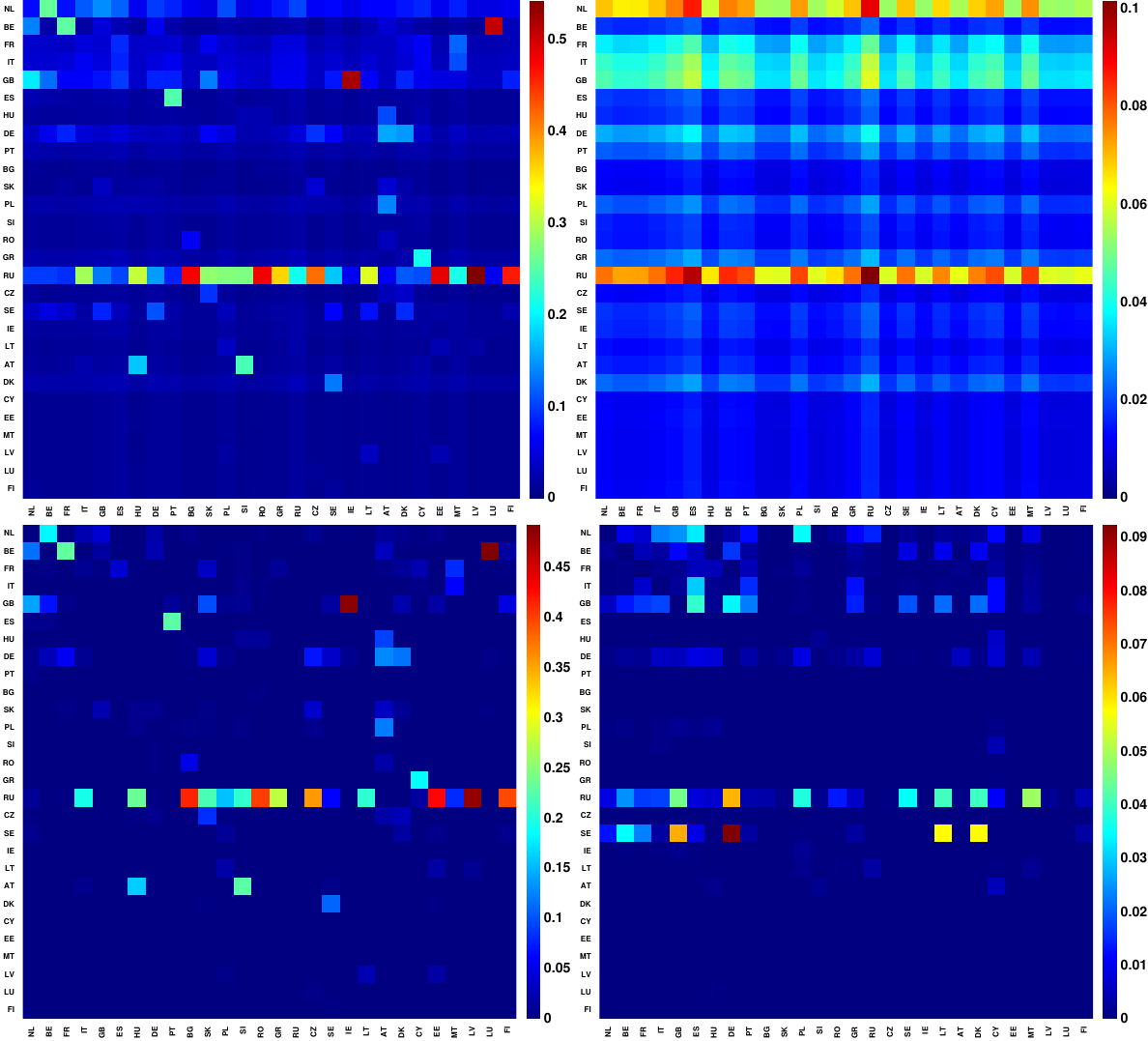}
\caption{
Left four panels: reduced Google matrix $\GR$ (top left) and its matrix components 
$\Gpr$ (top right), $\Grr$ (bottom left) and $\Gqrnd$ (bottom right) 
for the gas product (code $p=34$) exchanged among the 27 EU countries 
and Russia in 2016. Right four panels: the same as on the left but
for reduced Google matrix $\GR^*$ and its three matrix components
in the same order as on the left.
Here, the EU countries and RU are ordered as in
the PageRank column of Table~\ref{tab:gas16}.
}
\label{fig2}
\end{figure*}

Examples of $\GR$ and its 3 matrix components 
are shown in Fig.~\ref{fig1} for 
27 EU countries with code $p=33$ 
(27 nodes) plus petroleum of Russia, i.e., a total of 28 nodes for $\GR$  
(from the global network with $N=13 847$ nodes).
The same $\GR$ matrix but for gas from Russia
is shown in Fig.~\ref{fig2}. 

We discuss the properties of these $\GR$ matrix components 
shown in Figs.~\ref{fig1},~\ref{fig2}
in the next Section.
Here we only note that for selected countries
this $\GR$ matrix captures only
trade in petroleum (or gas).
This can be interesting in itself but
in this way we cannot obtain
the country balance and its sensitivity.
Thus we follow another approach.
We take 27 EU countries with all their products
(that gives us $27 \times 61 = 1647$ nodes) and we add
to these nodes the node of RU-petroleum.
In this way we obtain  $\GR$ matrix with 
the size of $N_r=1647+1=1648$ nodes
(from the total size of $G$ being $N=13847$).
In this $\GR$ matrix we have all 
direct and indirect links of all products of 27 EU countries
with petroleum of RU. In this $\GR$ matrix we can change the petroleum price
using the multiplier $(1+\delta)$ for links from
RU petroleum to other nodes 
with the renormalization of all 
matrix elements in this column to unity.
Then we obtain the probabilities $P_{pc}$ for all $27+1$ countries.
The same procedure is done for the CheiRank $\GR$ matrix
getting $P^*_{pc}$ and then the balance sensitivity $dB_c/d\delta$
of country (including all its products)
to Russian petroleum.
The same procedure is used to obtain the sensitivity to
Russian gas (or US or other country gas).
The sen\-si\-ti\-vi\-ty computed in this way
gives us the real sen\-si\-ti\-vi\-ty of country balance 
taking into account all direct and indirect links
present in the WTN.

\section{Results}
\label{sec:3}

Here we present the results for EU trade obtained with the reduced Google matrix
algorithm.

\subsection{Examples of reduced Google matrices $\GR$ and $\GR^*$}

In Fig.~\ref{fig1} we show the reduced Google matrices $\GR$ and $\GR^*$
and their  3 matrix components
for EU petroleum trade with Russia  2016.
The matrix size of selected nodes is $N_r=27+1=28$,
the direct and indirect link contributions from other network nodes
$N_s=N-N_r = 13847 - 28 = 13819$ are taken into account
by the REGOMAX algorithm. The nodes are ordered by the PageRank index
of countries given in Table~\ref{tab:petroleum16}.

We characterize the weight $\WR$, $\Wpr$,
$\Wrr$, $\Wqr$
of $\GR$ and its 3 matrix components $\Gpr$, $\Grr$, $\Gqr$
by the sum of all its elements divided by the matrix size $N_r$
($\Wqrnd$ for $\Gqrnd$).
By definition we have  $\WR=1$. It is usual for Wikipedia
networks that the weight $\Wpr \approx 0.95$ (see e.g. \cite{politwiki,wrwu2017})
is rather close to unity since $\Gpr$ is approximately composed 
from identical columns of PageRank vector, while the remaining 
weight of about $0.05$ is approximately equally distributed between
$\Wrr$ and $\Wqr$. We find that for the WTN the situation is different.
We have $\Wpr = 0.651568$, 
$\Wrr = 0.30849$, $\Wqr = 0.039942$ and $\Wqrnd = 0.036512$
 so that the weight of $\Wpr$ is significantly reduced
and the weight of $\Wrr$ is significantly larger than the weight
of $\Wqr$. We attribute this to the fact that
the global $S$ matrix of WTN contains many links
(about $~2000$ links per node for matrix elements
with amplitude being larger than $\sim10^{-4}$)
in contrast with the  very sparsed Wikipedia $S$
matrix. Hence, for WTN the importance of direct links is significantly higher.
For $\GR^*$ and its 3 matrix components we obtain the following weights
$\Wpr^* = 0.6051$, $\Wrr^* = 0.34379$, $\Wqr^* = 0.05111$ and $\Wqrnd^* = 0.047$
which are similar to the $\GR$ case.

In Fig.~\ref{fig1} (left 4 panels) we show $\GR$ matrix with its 3 matrix components for 
petroleum product (code $p=33$) trade of 27 EU countries with Russia.
For $\GR$ and $\Grr$ the dominant matrix elements correspond to trade flow
from Republic of Ireland (IE) to UK (GB).
Indeed, since UK and IE both have territories
on island of Ireland the trade flow between two countries is very high.
The next by the amplitude is the trade flow from
Denmark (DK) to Sweden (SE) both in $\GR$ and $\Grr$.
Among the indirect links in $\Gqr$ we find as the strongest the flow
from Portugal (PT) to Spain (ES)
and from Romania (RO) to Bulgaria (BG)
and Cyprus (CY) to Italy (IT).
However, the amplitude
of these transitions is relatively small.
In the matrix component $\Gpr$ the dominant transitions points to top PageRank countries NL, FR, DE, ES, GB, IT.
In all the matrix components the contribution of petroleum
from RU is not very pronounced.
We see the similar features for the petroleum
trade from US and SA shown in Figs.~\ref{figA1} and \ref{figA2}
of Appendix.  These results show that the contribution
of petroleum trade is masked by the active trade
between EU countries with other products.

The reduced Google matrix $\GR^*$ and its 3 matrix components
are presented in Fig.~\ref{fig1} (right 4 panels).
Here, we keep in mind that the flow directions have
been inverted to compute CheiRank probabilities. 
Thus to obtain the highest petroleum exports
from Russia we have to focus on the largest matrix elements
on the RU horizontal line. Contrarily to the $\GR$ case, here RU exports of petroleum clearly dominate the $\GR^*$ matrix and its 3 matrix components; this is mainly due to the fact that RU is the petroleum top exporter (see CheiRank and ExportRank in Table~\ref{tab:petroleum16}). From $\GR^*$ we observe that the strongest petroleum flows from RU point (in decreasing importance) to
Latvia (LV), Lithuania (LT), Finland (FI), BG, Poland (PL),   Estonia (EE), ... which are countries peripheral to RU. We also note non negligible petroleum flows from NL to Belgium (BE), and from BE to Luxembourg (LU).

\begin{figure*}[!ht]
\begin{center}
\includegraphics[width=\textwidth]{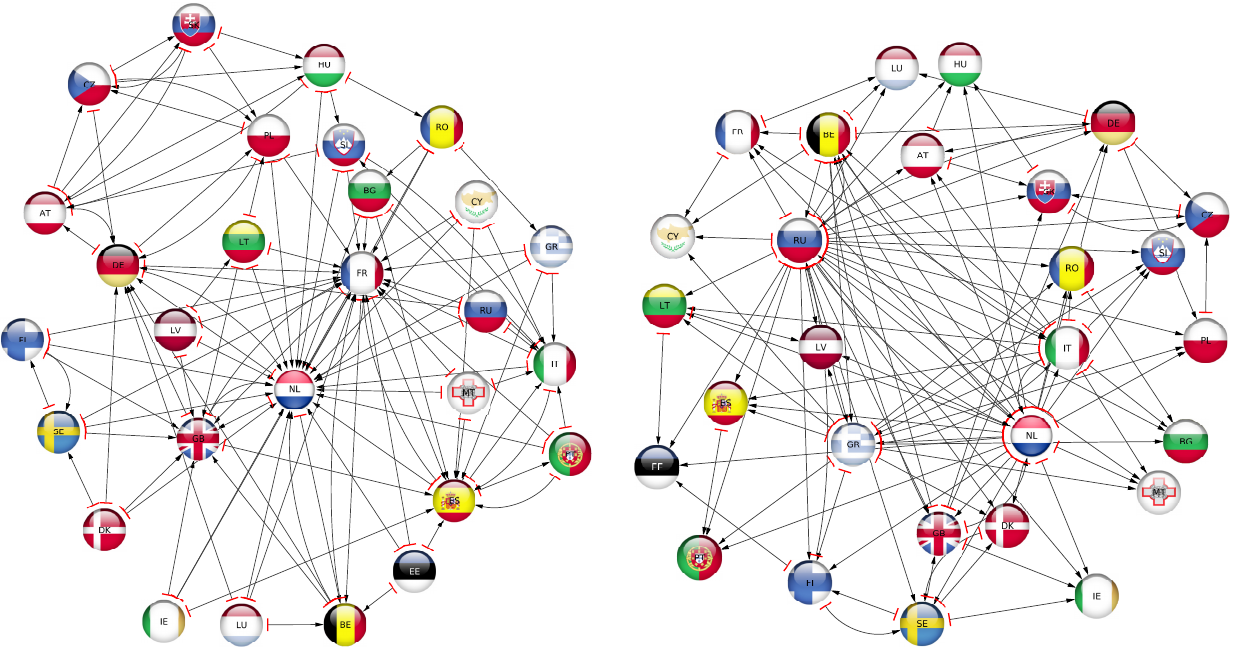}
\end{center}
\caption{
Network of petroleum import and network of 
petroleum export between EU countries and Russia in 2016. 
The EU and Russia petroleum reduced network is built from $\GR$ 
for import (left panel) and from $\GR^*$ for export (right panel). The network construction rule is the following: for each country $c$, 
we determine the 4 best petroleum importers from (exporters to) country $c$ according to $\GR$ ($\GR^*$). The directed links illustrate the petroleum flows.
}
\label{fig3}
\end{figure*}

Fig.~\ref{fig2} presents reduced Google matrices $\GR$ and $\GR^*$ for EU gas trade with Russia. The weights of the
$\GR$ and $\GR^*$ matrix components
$\WR=1$, $\Wpr=0.634069$, $\Wrr=0.308960$, $\Wqr=0.056971$ $(\Wqrnd=0.051085)$ and
$\WR^*=1$, $\Wpr^*=0.611761$, $\Wrr^*=0.322066$, $\Wqr^*=0.066173$ $(\Wqrnd^*=0.058111)$,
are similar to those of EU petroleum trade with Russia. In Fig.~\ref{fig2} (left 4 panels), the $\GR$ matrix gives the strongest gas import flows which are, by decreasing importance, from CY to IT, IE to GB, DK to SE, BE to FR, LT to PL, ES to PT, ... The $\Gpr$ matrix component shows that the main importer for gas in EU is NL. Indeed import flows toward NL are at least about one order of magnitude more important than toward the other EU states and in particular FR which is nonetheless the top importer according to ImportRank (see Table~\ref{tab:gas16}). For the case of gas trade between EU countries and RU, we note that the maximum matrix elements in $\Gqr$ have the same magnitude than the maximum matrix elements of the other matrix components. In particular, hidden indirect import flows from LU and PL toward Hungary (HU) are clearly visible from $\Gqr$. In Fig.~\ref{fig2} (right 4 panels), from $\GR^*$, the strongest gas export flows emanate mainly from RU toward (by decreasing importance) Latvia (LV), EE, FI, BG, RO, ... Besides this Russian gas export, the second and third most important gas export flows are in fact from GB to IE and from BE to LU. From $\Gpr^*$ we see that gas export flows from NL, which is the top EU gas exporter according to CheiRank (see Table~\ref{tab:gas16}), although weaker than the ones from RU are nonetheless of the same order of magnitude. Among EU countries and Russia, NL and RU compete for the best gas supplier. Although the weight of $\Gqr^*$ is weaker than the weight of the other matrix components, hidden indirect gas export flows can be seen in $\Gqr^*$ from, SE and RU, to DE.

\subsection{Network structure of petroleum and gas EU trade}

From $\GR$ and $\GR^*$ matrices shown in Fig.~\ref{fig1}, we are able to extract the network structure of the petroleum trade between EU countries and RU.
Fig.~\ref{fig3} left panel shows the petroleum import trade network between EU and RU. The top 6 EU economies by nominal GDP (i.e. DE, GB, FR, IT, ES, NL in 2016 \cite{wikigdpEU}) are the main petroleum importers, NL and FR being the more central. The performances of these economies are consequently correlated to their abilities to efficiently import petroleum. The four main direct and/or indirect EU gates for RU petroleum are DE, FR, NL, IT. We note closed loop petroleum exchange between (almost) neighboring countries, e.g. DE-AT, CZ-SK, DE-PL, AT-HU, AT-SK, PT-ES, ES-IT, SE-FI.
Fig.~\ref{fig3} right panel shows the petroleum export trade network between EU and RU. We clearly retrieve the fact that RU is the first petroleum supplier of EU and that NL is the top EU exporter of petroleum (see CheiRank and ExportRank in Table~\ref{tab:petroleum16}). From both of the petroleum trade networks shown in Fig.~\ref{fig3} we observe that NL constitutes the main European hub for petroleum exchanges. Secondary petroleum exporters are GR, IT, BE, GB, SE, and DE.

\begin{figure*}[!ht]
\begin{center}
\includegraphics[width=\textwidth]{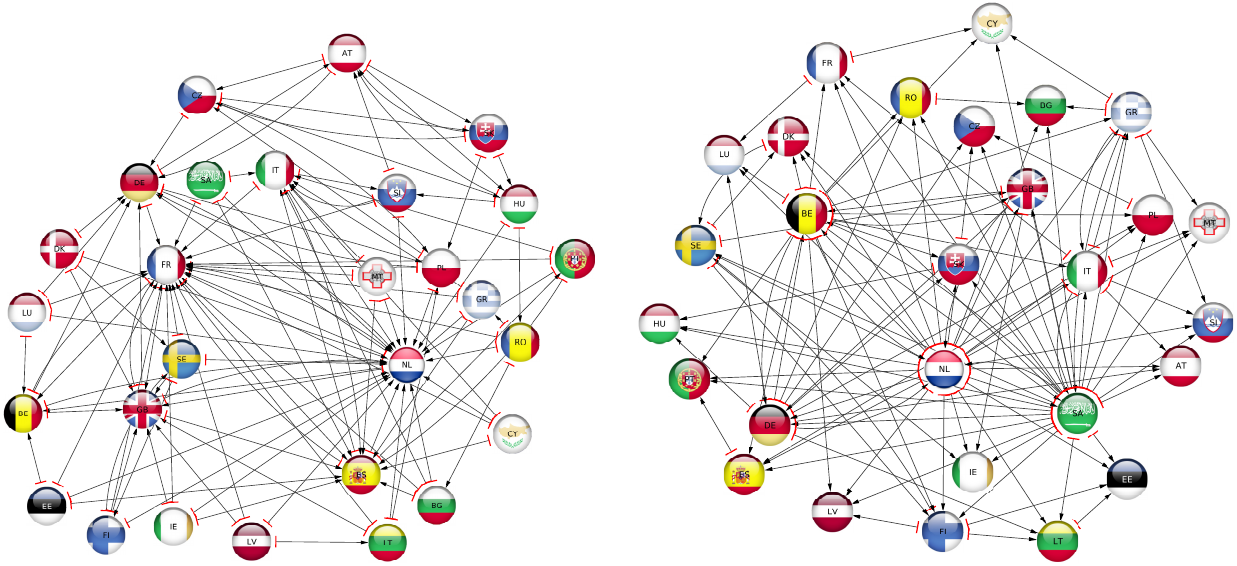}
\end{center}
\caption{
Network of petroleum import and network of 
petroleum export between EU countries and Saudi Arabia in 2016.
The EU and Arabia petroleum reduced network is built from $\GR$ 
for import (left panel) and from $\GR^*$ for export (right panel). 
The network construction rule is the following: for each country $c$, 
we determine the 4 best petroleum importers from (exporters to) country $c$ according to $\GR$ ($\GR^*$). 
The directed links illustrate the petroleum flows.
}
\label{fig4}
\end{figure*}

We also construct the reduced Google matrices $\GR$ and $\GR^*$ associated to petroleum import and export between EU countries and Saudi Arabia (SA).
Fig.~\ref{fig4} shows the petroleum trade network between EU countries and SA. The EU+SA petroleum import trade network (Fig.~\ref{fig4} left panel) is similar to the one obtain for EU+RU (see Fig.~\ref{fig3} left panel). This illustrate the robustness of the EU intramarket in regards to petroleum import. The main entrances in EU for SA petroleum are FR, IT, ES, and NL. Fig.~\ref{fig4} right panel shows that besides SA the main EU petroleum exporters are NL, BE, DE, IT, GR. From both of the EU+SA petroleum trade networks shown in Fig.~\ref{fig4} we observe a situation different from the EU+RU case (Fig.~\ref{fig3}) as not only NL but also DE, IT, GB, constitute each one a hub for petroleum exchanges. Although SA is the top petroleum exporter worldwide, RU is the main supplier for EU, this is the reason why trade networks with SA allows also to unveil secondary petroleum exchange hubs.

\begin{figure}[!t]
	\begin{center}
		\includegraphics[width=0.48\textwidth]{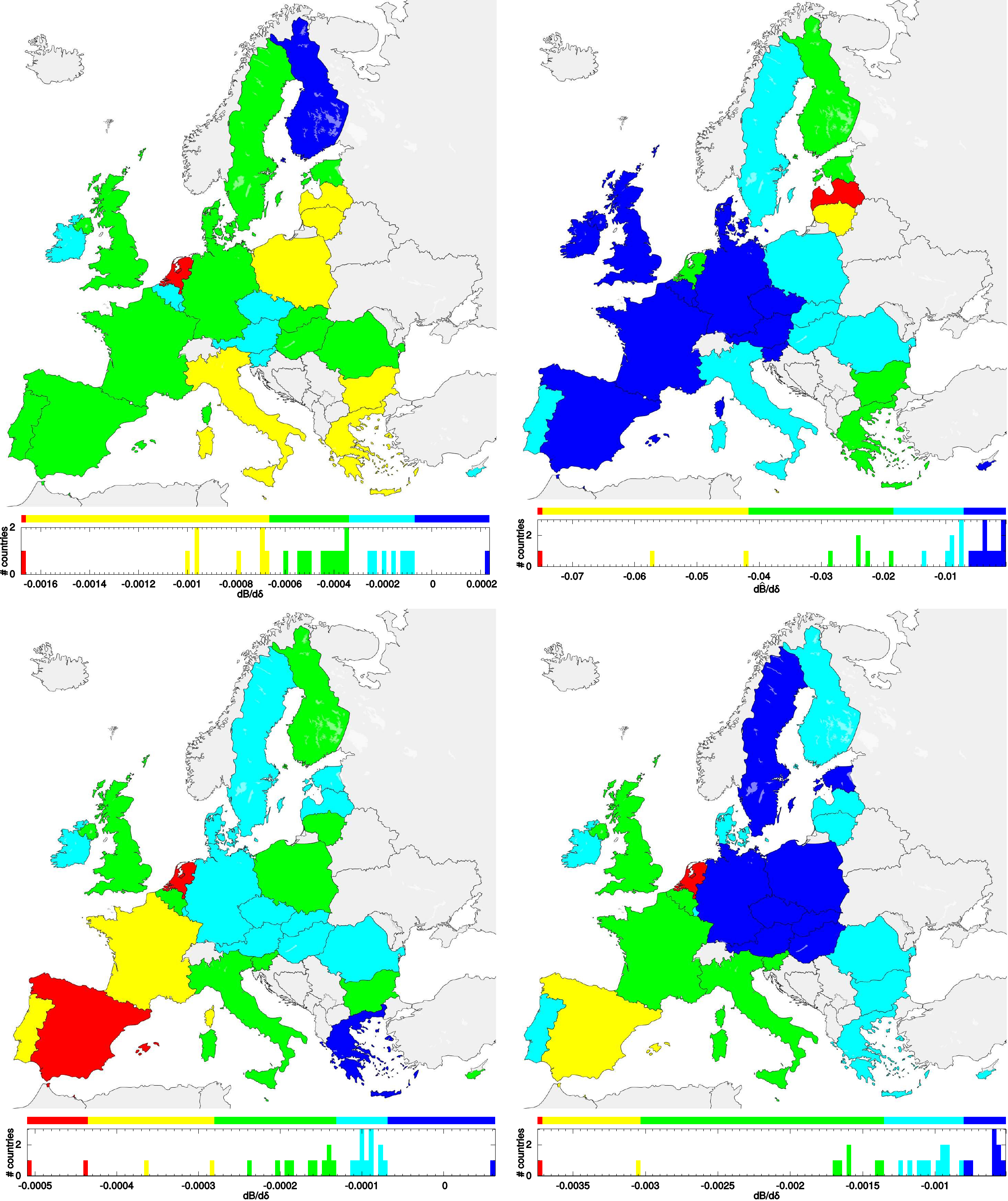}
	\end{center}
	\vglue -0.1cm
	\caption{
		EU countries balance derivative $dB_{c}/d\delta$ induced by an increase of petroleum price
		from Russia, Saudi Arabia, and United States in 2016. For each EU country $c$, 
		we compute the balance derivative $dB_{c}/d\delta$ induced by an infinitesimal 
		change of petroleum price  to EU country $c$ from Russia (top left), 
		from Saudi Arabia (bottom left), from United States (bottom right). 
		The balance derivative $d\hat B_{c}/d\delta$, 
		computed using ImportRank $\hat P_c$ and ExportRank $\hat P^*_c$, 
		induced by an increase of petroleum price from Russia to 
		EU country $c$ is shown in top right panel.
	}
	\label{fig5}
\end{figure}

\subsection{Sensitivity of EU to petroleum price}

Above we have considered the reduced Google matrices $\GR$ and $\GR^*$
with related networks only for petroleum or gas
flows of 27 EU countries plus Russia (or SA, US).
However, this approach does not capture the 
global influence of petroleum or gas trade on the \textit{all products} trade balance 
of a given EU country.
Therefore we extend our REGOMAX analysis taking into account
the matrix size $N_r = 1648$ for $\GR$ and $\GR^*$
(see Section 2.4). As the main characteristic we analyze the 
sensitivity of country global trade balance
in respect to small petroleum price increase 
(from unit price $1$ to price $1+\delta$)
expressed by the derivative $d B_c/ d \delta$.
As described in section~\ref{subs:sensitivity}
we express the country $c$ balance $B_c$ via CheiRank $P^*_c$ and PageRank $P_c$ probabilities
and also via ExportRank $\hat P_c^*$ and ImportRank $\hat P_c$ probabilities.

In Fig.~\ref{fig5} we present the sensitivity $d B_c/d \delta$,
shown by color, 
to petroleum trade with RU, US and SA on 
the EU political map  for year 2016.
The sensitivity to petroleum from Russia is shown in Fig.~\ref{fig5}
top left panel. We see that the strongest negative 
effect is produced on NL
which is on at the top PageRank position (see Table~\ref{tab:petroleum16})
due to its strong maritime relations which bring a lot of petroleum to NL and then
redistributed to other EU countries.
The next most sensitive EU countries are Italy (IT),
Greece (GR), Bulgaria (BG), Poland (PL), Lithuania (LT) and Latvia (LV).
We note that here the sensitivity $d B_c/d \delta$
is defined via CheiRank and PageRank probabilities
taking into account the multiplicity of WTN links.
The result is very different (see Fig.~\ref{fig5} top right panel) if 
the sensitivity $d \hat B_c/d \delta$ is defined by ExportRank and ImportRank 
probabilities, which are usually used in economy for
the trade analysis. This crude Export-Import analysis gives the most 
strong negative sensitivity for Latvia (LV). The next is Lithuania (LT)
which as LV keeps close trade relations with RU being ex-USSR republics.
Moreover the Export-Import analysis gives a rigid component of Western EU countries almost not sensitive to RU petroleum.
The drastic global difference between REGOMAX analysis and 
the simple standard Import-Export analysis is that the first considers 
the multilateral cascade of direct or indirect trades between two countries and 
the second only considers the direct bilateral trade between two countries.
We consider
that the REGOMAX algorithm provides 
much more detailed and realistic information
on sensitivity to petroleum
price compared to the usual Export-Import consideration.
We attribute this advantage of REGOMAX analysis to
its deep mathematical properties that allows
to take into account all direct and hidden links between
selected nodes of the WTN.
Due to these reasons below we focus mainly on results obtained with the REGOMAX analysis.

The sensitivity of EU to petroleum price
from SA and US are shown in the bottom panels of 
Fig.~\ref{fig5}. For SA the most sensitive countries are Spain (ES) and NL
while for US the most sensitive is NL.
However, for EU the sensitivity to petroleum of SA
is by a factor 3-4 smaller than for those from RU.
Thus the sensitivity of Germany to Russian petroleum is
by a factor 5 stronger than of SA petroleum.
In contrast the maximum EU sensitivity to US petroleum
is by a factor 2 stronger than to those of RU.
The sensitivity of Germany is comparable for US and RU.
Let us note that GR is not affected and even benefit from SA petroleum price increase. 
The same for FI benefiting from RU petroleum price increase.
Also we observe a rigid component of Eastern EU countries from Sweden to Greece and 
from Baltic countries to Germany which are almost insensitive to US petroleum (Fig.~\ref{fig5} bottom right panel).

\begin{figure*}[!ht]
\begin{center}
\includegraphics[width=\textwidth]{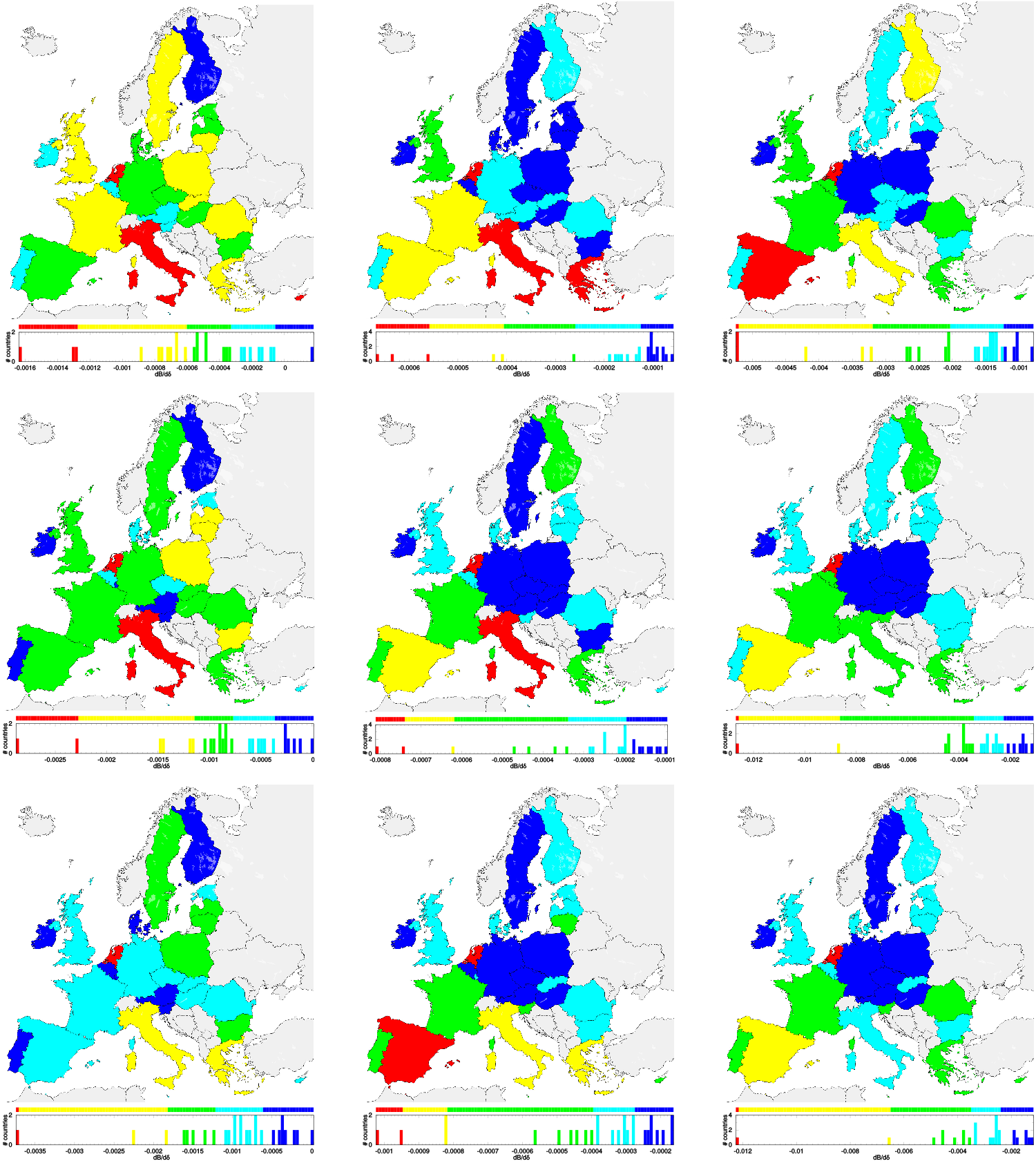}
\end{center}
\vglue -0.1cm
\caption{
EU countries balance derivative $dB_{c}/d\delta$ induced by 
an increase of petroleum price from Russia (left column), 
Saudi Arabia (middle column), and United States (right column), 
for 2004 (top row), 2008 (middle row), and 2012 (bottom row).
}
\label{fig6}
\end{figure*}

The time evolution of EU sensitivity to petroleum from RU, SA, and US
is shown in Fig.~\ref{fig6} for years 2004, 2008, 2012
(for year 2016 see previous Fig.~\ref{fig5}).
For RU petroleum
the most sensitive country are Netherlands (NL), Italy (IT), Cyprus (CY) in
2004, NL and IT in 2008 and NL in 2012 and 2016.
Also the maximal negative sensitivity is changing from
$-0.0016$ in 2004 to $-0.0029$ in 2008,
$-0.0037$ in 2012 and
$-0.0017$ in 2016.
From these maximal sensitivities and also from the distribution 
of sensitivities among EU countries, we observe an overall increase of 
the balance trade sensitivity to RU petroleum until 2012, then we remark that 
EU trade sensitivities in 2016 decreases being comparable back to those in 2004.
We attribute this to a significant drop of petroleum 
price happened in the world after the financial crisis of 2007-2008.
A similar tendency is visible for SA and US petroleum sensitivity.

\begin{figure}[!ht]
\begin{center}
\includegraphics[width=0.48\textwidth]{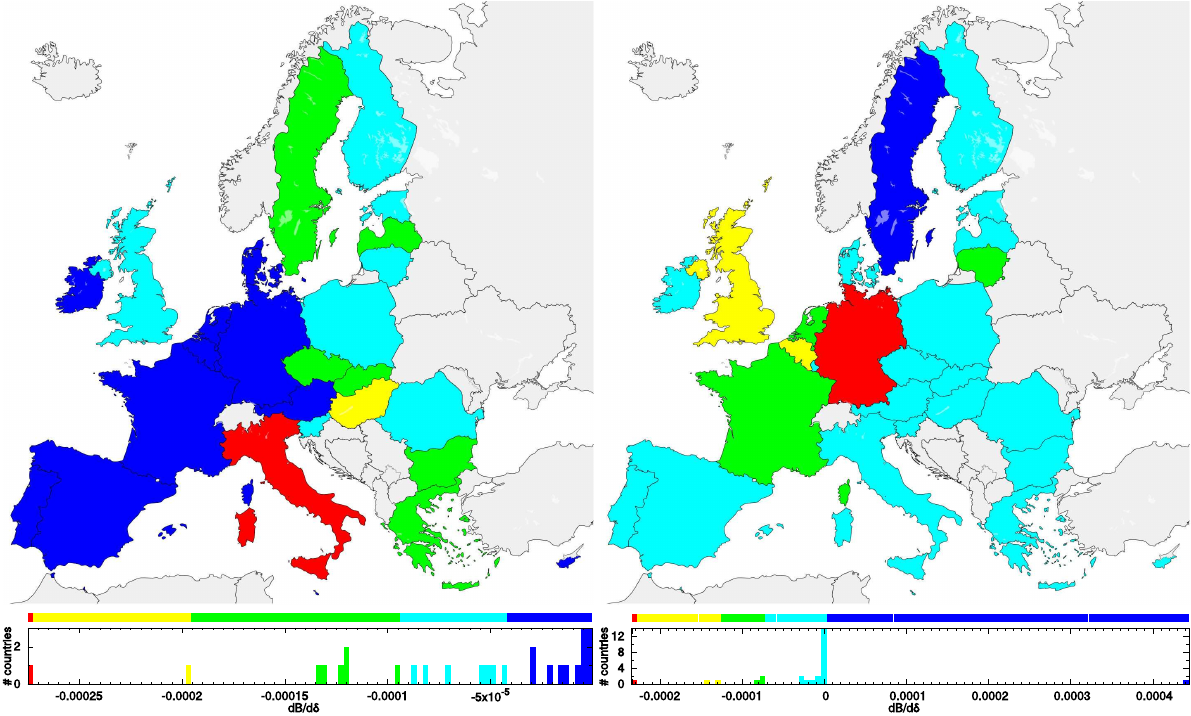}
\end{center}
\vglue -0.1cm
\caption{
EU countries balance derivative $dB_{c}/d\delta$ induced by an increase 
of gas price from Russia (left panel) and Norway (right panel) in 2016.
}
\label{fig7}
\end{figure}

For SA petroleum the most
sensitive countries are
NL, IT and Greece (GR) in 2004,
NL, IT in 2008, NL, Spain (ES) in 2012
and  2016 with the maximal negative sensitivity
changing from $-0.0006$ in 2004, $-0.0008$ in 2008,
$-0.001$ in 2012 and $-0.0005$ in 2016.
As in the RU case, trade balance sensitivities of EU countries 
to SA petroleum increases until 2012 and then decreases in 2016 to attain values comparable to year 2004.

For US petroleum the most
sensitive countries are
NL and ES in 2004 and NL in 2008, 2012 and 2016
with the maximal negative sensitivity changing from
$-0.0052$ in 2004, $-0.0127$ in 2008,
$-0.0122$ in 2012 and $-0.0037$ in 2016. 

Globally, the ancient USSR satellites and more globally central EU economies 
are less affected by the increase of petroleum from US or SA.
We also observe that due to NL central position in petroleum import and export for EU, 
the performance of NL economy is the most affected by petroleum price increases. 
This strong indirect feature is absolutely not captured 
by the standard Import-Export analysis picture (see e.g. Fig.~\ref{fig5} top right panel).
In global we see that EU countries are more sensitive to US petroleum
that is by a factor 2-3 stronger comparing to those of RU.
We relate this to the fact that US is the world top PageRank country
so that it has more global world influence on other countries.

\subsection{Sensitivity of EU to gas price}

We present in Fig.~\ref{fig7} EU trade balance sensitivity to gas from RU and Norway (NO) in 2016.
EU sensitivities to RU gas is one order of magnitude weaker for gas than petroleum. 
The price increase of the RU gas mainly affects Italy (IT) 
while  other Western EU countries being relatively not sensitive. 
Again RU neighboring countries are the most sensitives to RU gas import. 
The most sensitive EU economies to Norwegian gas are DE economy 
(and to a lesser extent GB and BE economies) which would be affected by NO gas price increase 
and SE economy which would benefit from it. The positive balance trade sensitivity 
for SE is certainly due to the entanglement of NO-SE economies. The others economies 
are insensitive to NO gas (see peak of fourteen EU countries with balance trade 
around 0 in Fig.~\ref{fig7} right panel).

\begin{figure}[!t]
\begin{center}
\includegraphics[width=0.48\textwidth]{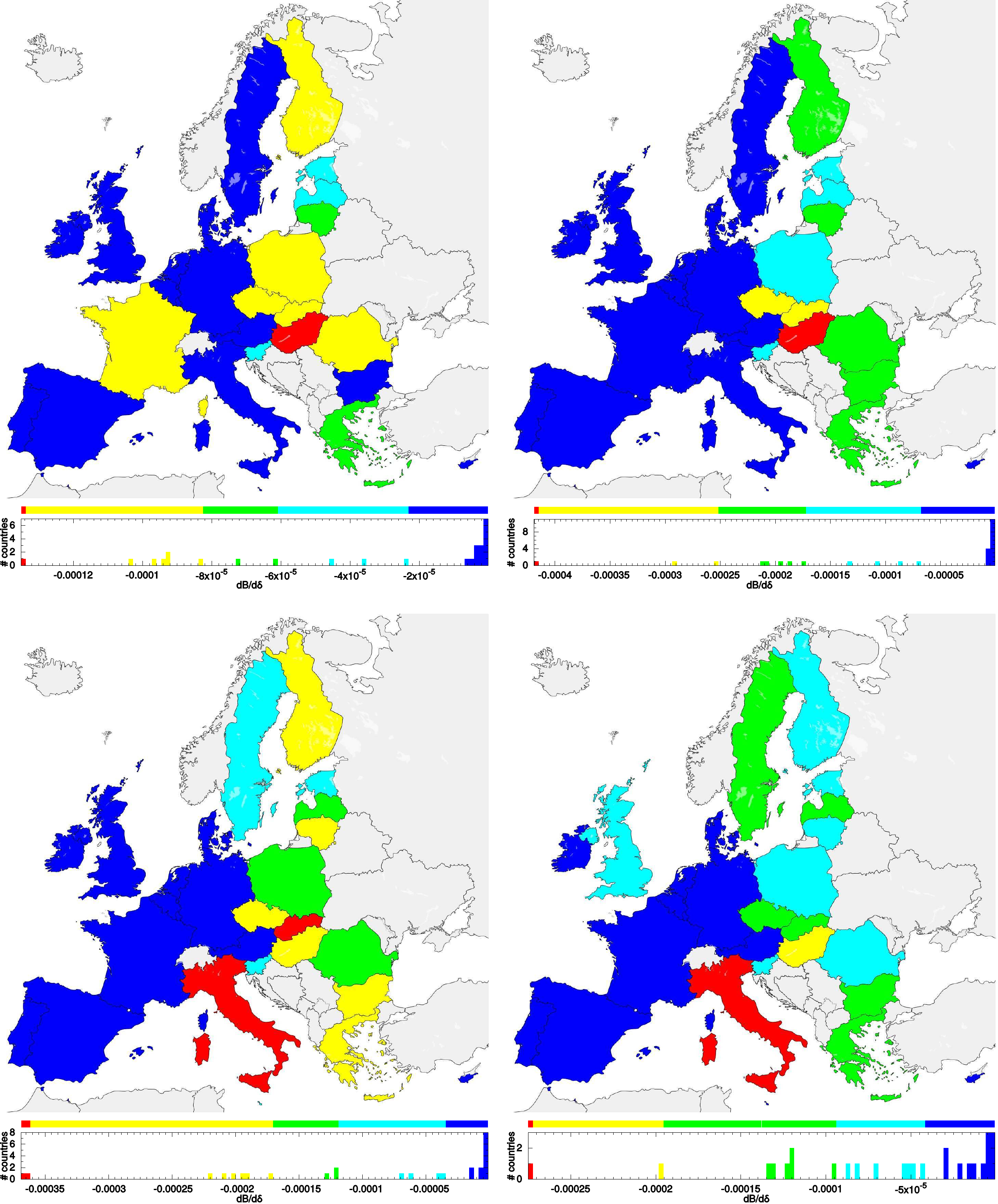}
\end{center}
\vglue -0.1cm
\caption{
EU countries balance derivative $dB_{c}/d\delta$ induced by 
an increase of gas price from Russia for 2004 (top left), 2008 (top right), 
2012 (bottom left) and 2016 (bottom right).
}
\label{fig8}
\end{figure}

\begin{figure}[!t]
\begin{center}
\includegraphics[width=0.48\textwidth]{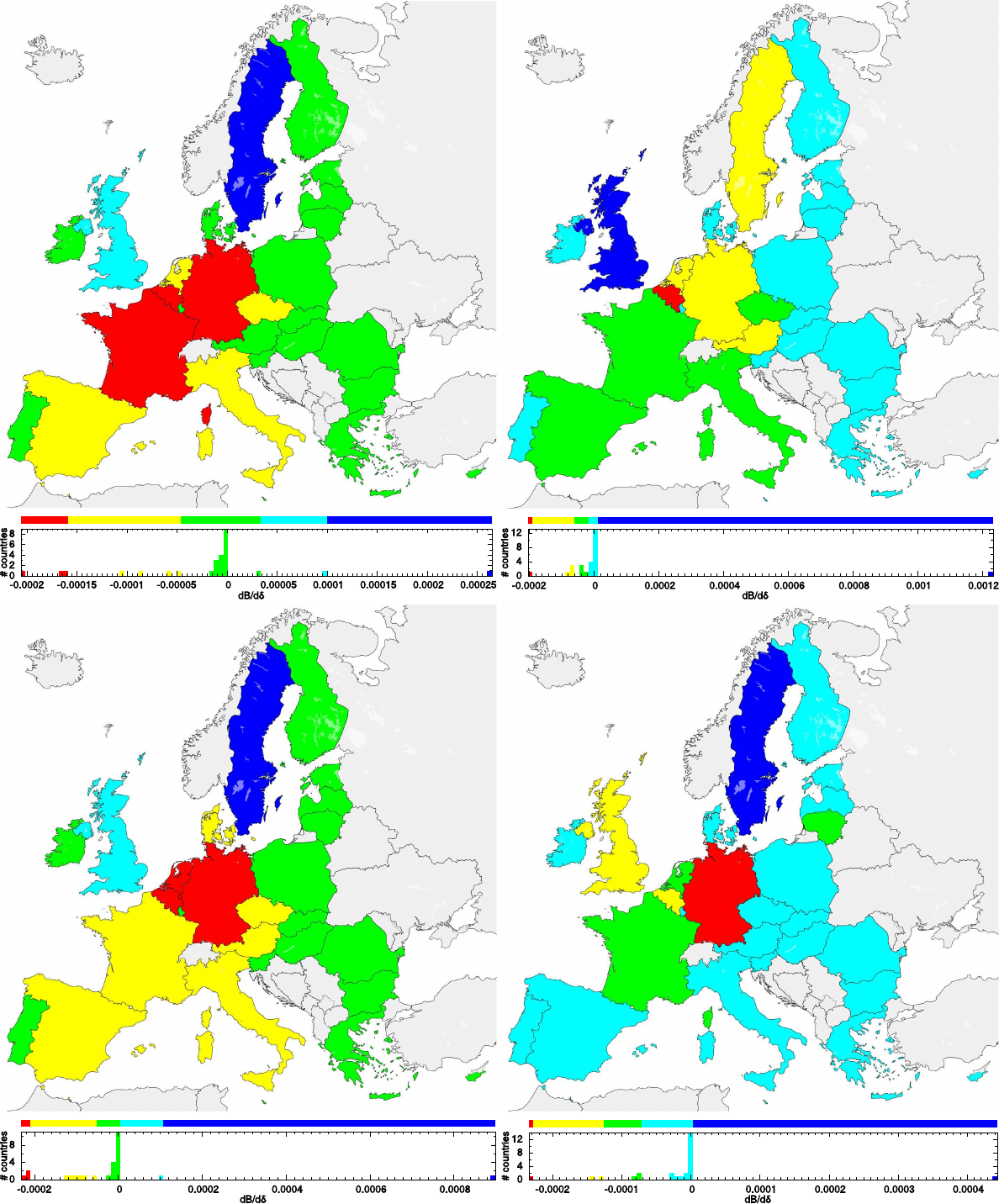}
\end{center}
\vglue -0.1cm
\caption{
EU countries balance derivative $dB_{c}/d\delta$ 
induced by an increase of gas price
from Norway for 2004 (top left), 2008 (top right), 
2012 (bottom left) and 2016 (bottom right).
}
\label{fig9}
\end{figure}

Figs.~\ref{fig8} and \ref{fig9} show from 2004 to 2016 time evolution of 
EU economies trade balance sensitivity to RU and NO gas. 
In Fig.~\ref{fig8} we observe that during this time period the Western EU 
bloc from PT to DE remained insensitive to RU gas with the exception of IT economy 
which became the most affected since 2012 
(also FR economy were temporarily sensitive to RU gas around 2004).
In Fig.~\ref{fig9} we observe that during the same period EU east end 
countries are insensitive to NO gas. The most affected countries by NO gas 
volume import and/or price increase are FR, BE, DE in 2004, BE in 2008, 
BE, NL, DE in 2012 and DE in 2016. SE economy always benefit from volume 
increase of NO gas excepting in 2008, at that time 
SE was relatively affected and GB was benefiting from NO gas.

\section{Discussion}

In this work we  developed the reduced Google matrix (REGOMAX) 
analysis of the multiproduct 
world trade network with a specific accent to sensitivity
of EU country trade balance to petroleum and gas prices
from Russia, USA, Saudi Arabia and Norway.
In particular we observe that, during the 2004-2016 time period, most of the EU countries 
are sensitive to price increase of Russian petroleum and petroleum products. 
The situation is different for Saudi Arabia and US petroleum price influence 
for which east and central EU countries are relatively less affected. 
The Netherlands, which is the best EU petroleum importer and exporter, 
is during this time period the most affected EU country by the price increase of 
either Russia, Saudi Arabia, or USA.
The influence of Russian gas is mostly exerted to Eastern EU countries among 
which ancient USSR satellites, Western EU countries being insensitive with the exception of Italy.
Although Norway is the second gas supplier for EU, the Norway price increase 
influences only few EU countries, affecting Germany during the whole 2004-2016 time period, 
France in 2004, Belgium in 2004 and 2012, and The Netherlands in 2012, but 
benefiting to Sweden (excepting around 2008).

We show that comparing to the usual export-import consideration
this REGOMAX approach takes into account
the cascade of chain influence propagation via all nontrivial
pathways of trade relations between countries. Due to this feature
this approach is
more powerful compared
to only nearby link analysis considered in the import-export approach.
Thus the REGOMAX method allows to recover indirect influence 
of petroleum or gas price from a specific country on EU trade.
We argue that the further investigation of 
such indirect influence will play an important role 
in petroleum or gas crisis contamination
propagation in EU trade.

\section*{Acknowledgments}
We thank the representatives of UN COMTRADE \cite{comtrade}
for providing us with the friendly access to this database.
This work was supported in part by the Pogramme Investissements
d'Avenir ANR-11-IDEX-0002-02, reference ANR-10-LABX-0037-NEXT 
(project THETRACOM);
it was granted access to the HPC resources of 
CALMIP (Toulouse) under the allocation 2017-P0110.
This work was also supported in part by the Programme Investissements
d'Avenir ANR-15-IDEX-0003, ISITE-BFC (project GNETWORKS) and 
by Bourgogne Franche-Comt\'e region (project APEX).

\section*{Appendix}
\setcounter{figure}{0}
\renewcommand{\thefigure}{A\arabic{figure}}

\begin{figure*}[h]
\centering
\includegraphics[width=0.99\columnwidth]{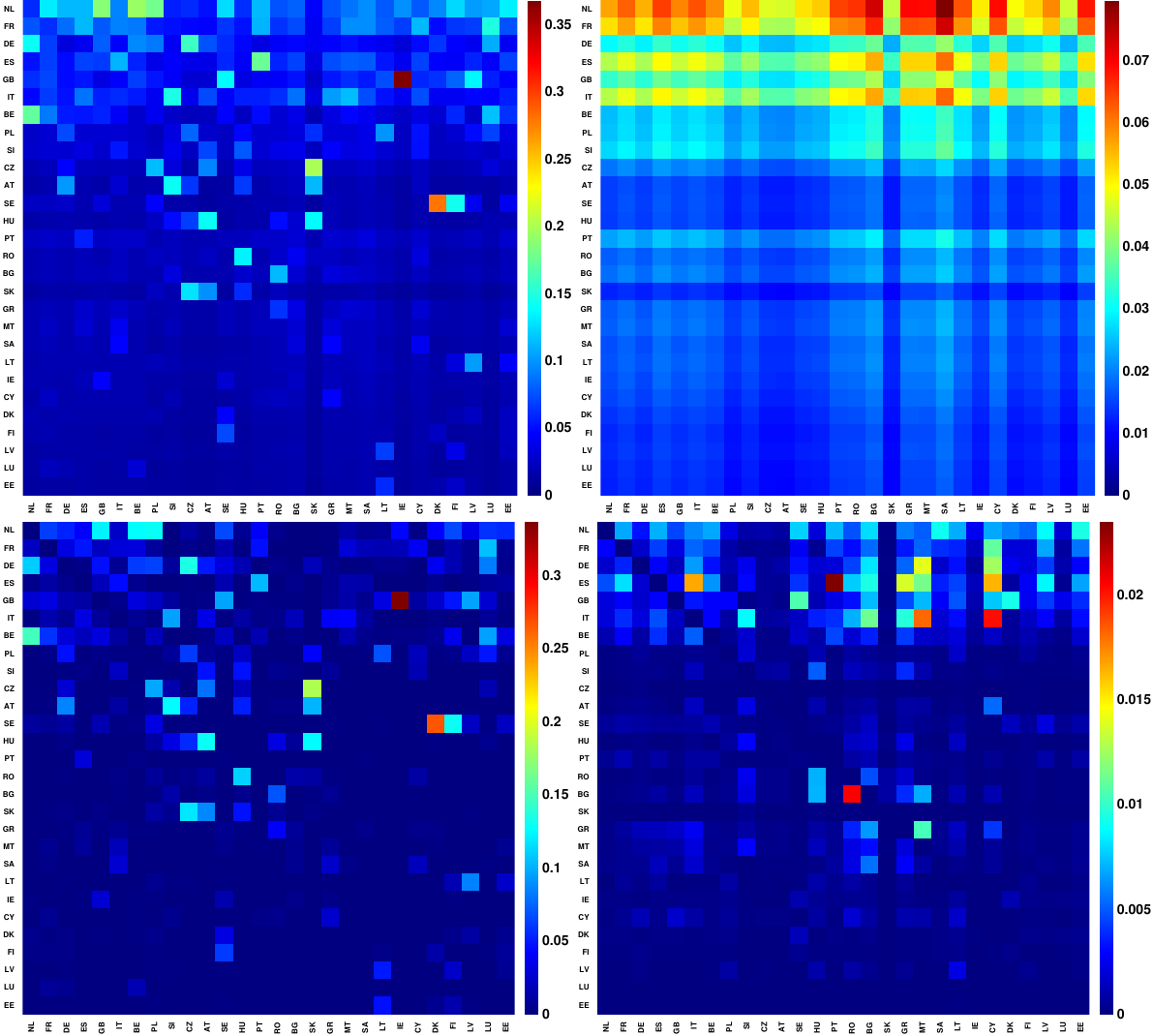}\hfill
\includegraphics[width=0.99\columnwidth]{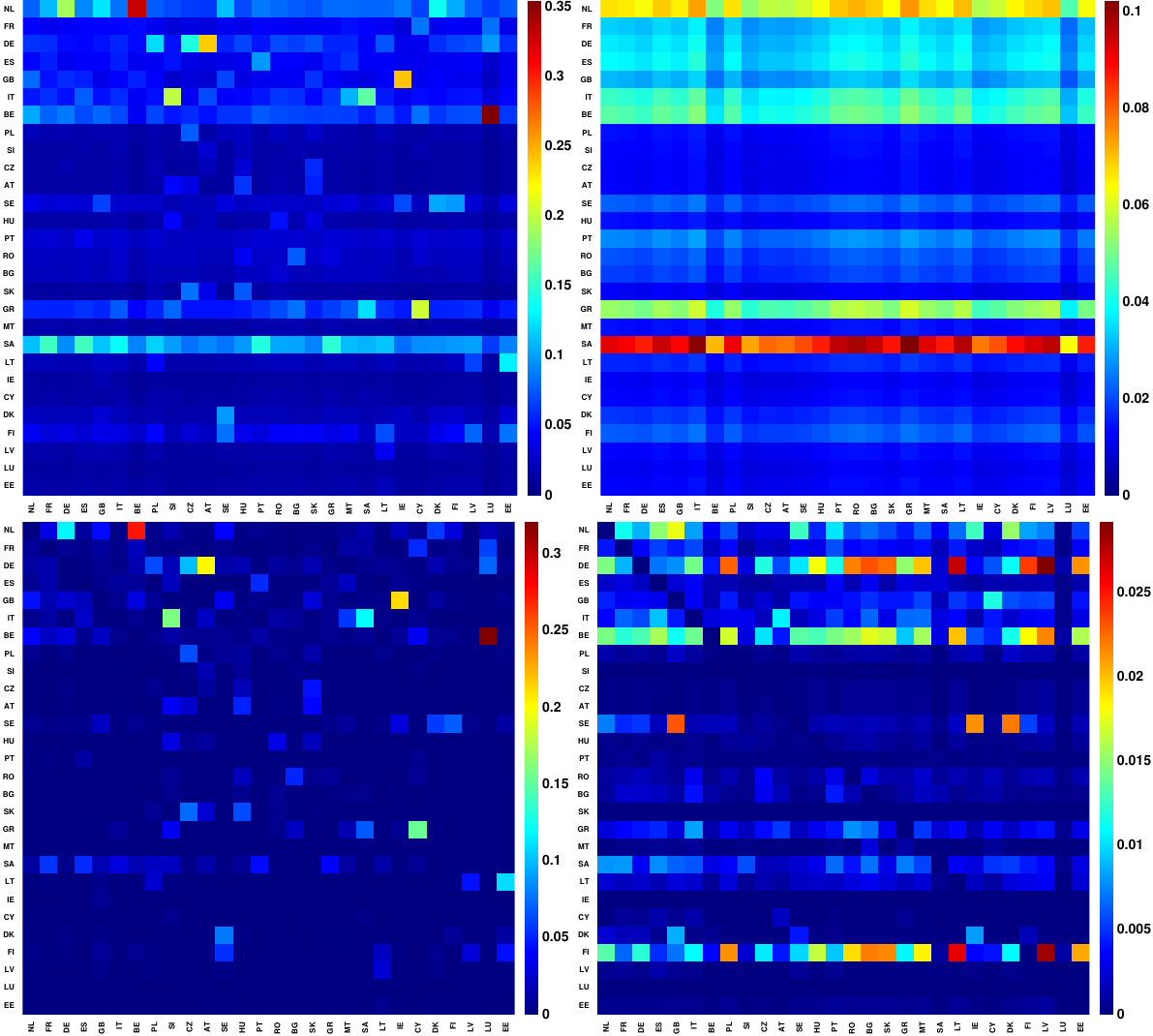}
\caption{
Left four panels: reduced Google matrix $\GR$ (top left) and its matrix components 
$\Gpr$ (top right), $\Grr$ (bottom left) and $\Gqrnd$ (bottom right) 
for the petroleum product (code $p=33$) exchanged among the 27 EU countries 
and Saudi Arabia in 2016. Right four panels: the same as on the left but
for reduced Google matrix $\GR^*$ and its three matrix components
in the same order as on the left.
Here, the EU countries and SA are ordered as in
the PageRank column of Table~\ref{tab:petroleum16}.
}
\label{figA1}
\end{figure*}

\begin{figure*}[h]
\centering
\includegraphics[width=0.99\columnwidth]{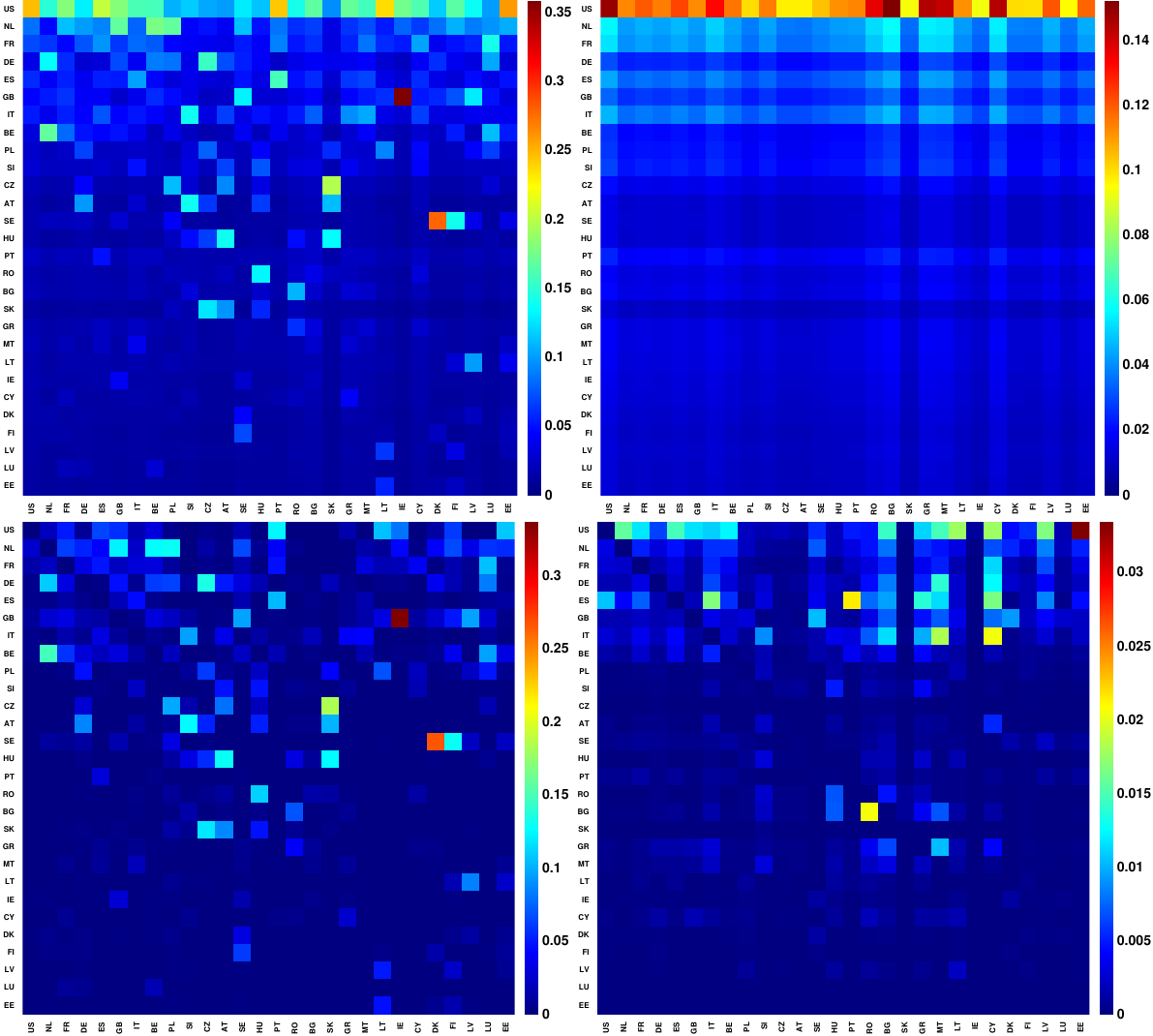}\hfill
\includegraphics[width=0.99\columnwidth]{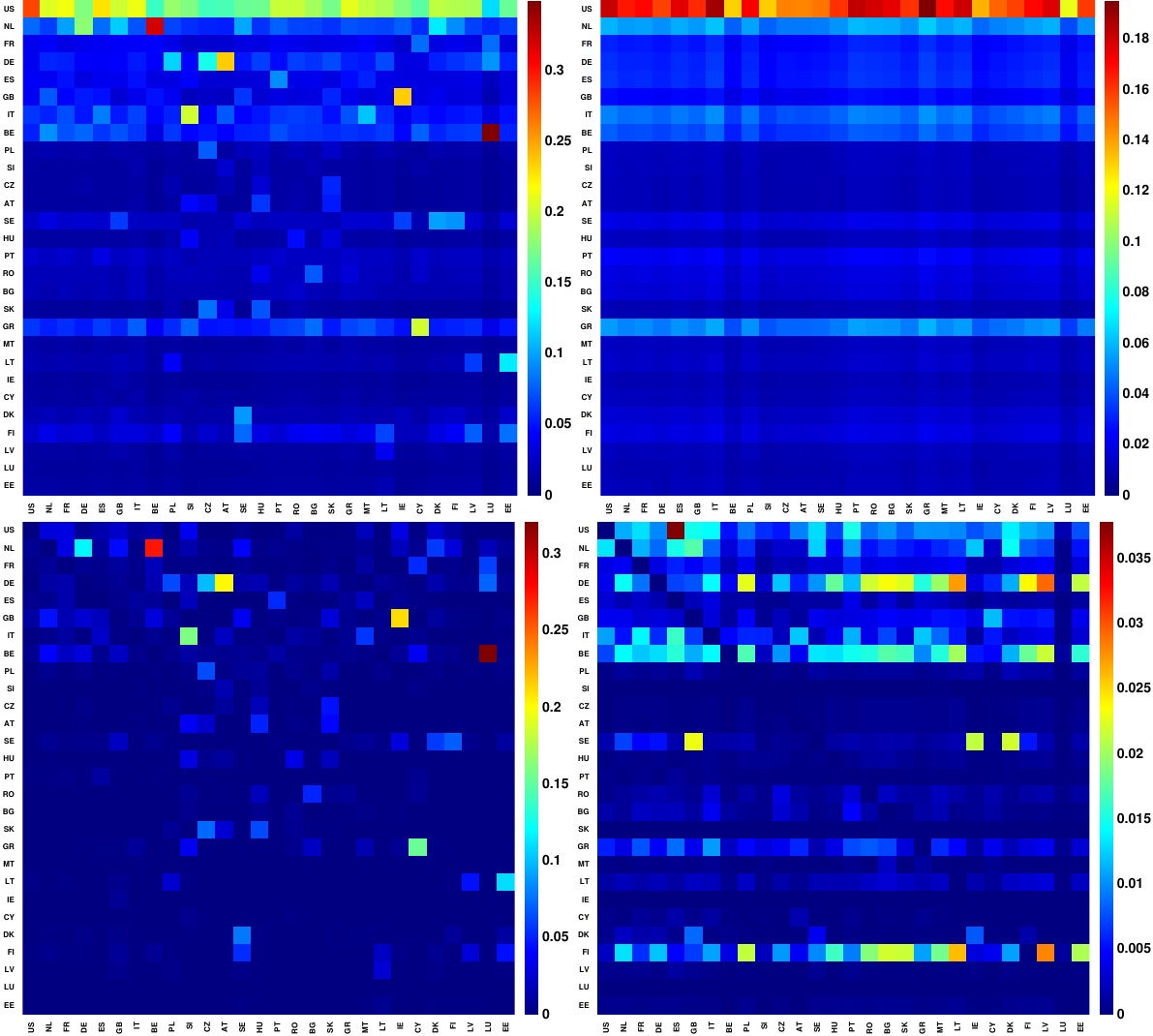}
\caption{
Left four panels: reduced Google matrix $\GR$ (top left) and its matrix components 
$\Gpr$ (top right), $\Grr$ (bottom left) and $\Gqrnd$ (bottom right) 
for the petroleum product (code $p=33$) exchanged among the 27 EU countries 
and USA in 2016. Right four panels: the same as on the left but
for reduced Google matrix $\GR^*$ and its three matrix components
in the same order as on the left.
Here, the EU countries and US are ordered as in
the PageRank column of Table~\ref{tab:petroleum16}.
}
\label{figA2}
\end{figure*}

Here we present some additional figures of
the reduced Google matrix analysis of EU trade. Fig.~\ref{figA1} shows the reduced Google matrices $\GR$ and $\GR^*$ for petroleum product associated to 27 EU countries and Saudi Arabia. Fig.~\ref{figA2} shows the reduced Google matrices $\GR$ and $\GR^*$ for petroleum product associated to 27 EU countries and US.

\onecolumn

\end{document}